\documentclass[10pt]{iopart}
\expandafter\let\csname equation*\endcsname\relax

\expandafter\let\csname endequation*\endcsname\relax
\usepackage{amsmath, amsthm, amssymb, amsfonts}
\usepackage{graphicx}
\usepackage{caption}
\usepackage{subcaption}
\usepackage{tabularx}
\usepackage{xfrac}
\usepackage{iopams}
\usepackage{hyperref}
\hypersetup{colorlinks,
citecolor=blue, urlcolor=blue, linkcolor=blue}
\usepackage{upgreek}
\usepackage[greek,english]{babel}
\usepackage[utf8]{inputenc}
\usepackage{orcidlink}
\usepackage{multirow}

\begin{document}
\title[Charged black holes in Weyl conformal gravity]{Charged black holes in Weyl conformal gravity}

\author{Reinosuke Kusano$^1$\orcidlink{0000-0001-7569-7197},
Miguel Yulo Asuncion$^{1,2,3}$\orcidlink{0009-0003-5075-3107}, Keith Horne$^1$\orcidlink{0000-0003-1728-0304}}
\address{$^1$SUPA Physics and Astronomy, North Haugh, University of St\,Andrews, KY16\,9SS, Scotland,
United Kingdom}
\address{$^2$Nottingham Centre of Gravity, Nottingham NG7 2RD, United Kingdom}
\address{$^3$School of Mathematical Sciences, University of Nottingham,
University Park, Nottingham NG7 2RD, United Kingdom}
\ead{rk77@st-andrews.ac.uk}
\vspace{10pt}
\begin{indented}
\item[]December 2025 
\end{indented}

\begin{abstract}
We present a parametric study of the spacetime structures obtainable in Weyl conformal gravity's dyonic Reissner-Nordstr\"{o}m solution. We derive expressions for photon sphere radii and horizons for this metric in terms of the conformal gravity parameters, from which we then determine analytic formulae for extremal limits and Hawking temperatures. Due to the surprising lack of the inverse quadratic $1/r^2$ term in this fourth-order metric, there is no guarantee for the innermost horizon of a black hole spacetime to be a Cauchy horizon, which is in direct contrast to the corresponding metric in general relativity. For example, for certain parameter values, a ``nested black hole'' is seen to exist; in such a spacetime, we find a Cauchy horizon trapped between two event horizons, which is not a structure known to be obtainable in standard general relativity. In addition to such exotic spacetimes, we also find a critical value for the electric and magnetic charges, at which the stable and unstable photon spheres of the metric merge, and we obtain extremal limits where three horizons collide. 

\vspace{1pc}
\end{abstract}

\submitto{\CQG}
%
%
%
\section{Introduction} 

Our modern understandings of space, time, and gravity are shaped by general relativity~(GR), the theory conceptualised by 
Einstein in 1915~\cite{Einstein:1915ca}. GR is encapsulated in the Einstein field equations:
\begin{equation}
    G_{\mu\nu} = 8\, \uppi \, T_{\mu\nu}.
\end{equation}
Gravity arises from spacetime curvature as quantified by
the Einstein tensor $G_{\mu\nu}$, while mass-energy distributions producing the curvature are encoded in the energy-momentum tensor $T_{\mu\nu}$. 
We here use geometrised units $(G = c = \hbar = 1)$ and a metric $g_{\mu \nu}$ signature of $(-, +, +, +)$.

GR enjoys many successes on solar system spatial scales, but tension arises on galactic scales where GR needs dark matter halos to account for observed flat galactic rotation curves~\cite{flat, flat2} and anomalous galaxy cluster dynamics~\cite{zwicky, roodcluster}. The discovery of the accelerating expansion of the universe~\cite{riessexpansion,perlmutter1999} then forces GR to embrace a repulsive cosmological constant or vacuum energy now known as dark energy.
GR also faces the vacuum energy problem, wherein cosmological observations and predictions of quantum field theory differ by around 120 orders of magnitude~\cite{weinberg}. 
While extensive work has gone into modifying GR on astronomical scales~\cite{brans1961, buchdal1970, bekenstein2004} as well as applying it to quantum regimes~\cite{penrose_1973,
scherk1974, yoneya1974, roverlli1990, finster1999_1, finster1999_2}, the theory 
overall leaves much to be desired. 

Weyl conformal gravity (CG), first proposed by Weyl~\cite{Weyl1918} and Bach~\cite{Bach1921} at the start of the 20th century, and further developed by Mannheim and Kazanas in 1989~\cite{mannheim1989}, is an alternative theory of gravitation with potential to relax the aforementioned astrophysical~\cite{horne_2006, Mannheim2012Rot, Mannheim2013Rot, mannheim_2022} and cosmological~\cite{mannheimgammavalue, VarieschiParameters1, VarieschiCosmo, mannheim2011} tensions without invoking the dark sector. In addition to the coordinate $g_{\mu \nu}(x) \rightarrow g'_{\mu \nu} (x')$, Lorentz $x^\mu \rightarrow \Lambda^\mu_\nu \ x^\nu$, and diffeomorphism invariances of GR~\cite{Einstein:1915ca}, CG possesses a further invariance to local conformal transformations $g_{\mu \nu}(x) \rightarrow \widetilde{g}_{\mu \nu}(x) = \Omega^2(x) g_{\mu 
\nu}(x)$, where $\Omega(x)$ is a stretching factor. 

The CG gravitational action is built from the conformal Weyl tensor $C_{\lambda \mu \nu \kappa}$, which is the Riemann tensor with all its traces removed: 
\begin{align}\label{eq:Weyl_tensor}
C_{\lambda \mu \nu \kappa} = R_{\lambda \mu \nu \kappa} -& \dfrac{1}{2}(g_{\lambda\nu} R_{\mu\kappa} - g_{\lambda\kappa}R_{\mu\nu} \notag - g_{\mu\nu}R_{\lambda\kappa}+g_{\mu\kappa}R_{\lambda\nu}) \notag \\&+ \dfrac{1}{6}R(g_{\lambda\nu}g_{\mu\kappa} - g_{\lambda\kappa}g_{\mu\nu}),
\end{align}
where $R_{\lambda \mu \nu \kappa}$ and $R_{\mu\nu}$ are the Riemann and Ricci tensors respectively. 
Squaring $C_{\lambda \mu \nu \kappa}$ then generates a scalar with dimensions of length$^{-4}$ suitable for the Lagrangian density
in the gravitational Weyl action: 
\begin{equation}\label{eq:Weyl_action}
    I_{\rm W} = -\alpha_g \int \text{d}^4 x \,\sqrt{-g}\, C_{\lambda \mu \nu \kappa}\,C^{\lambda \mu \nu \kappa}\ .
\end{equation}
Here the metric determinant is $g=\mathrm{det}(g_{\mu\nu})$, and $\alpha_g$ is a dimensionless gravitational coupling constant.
$\alpha_g<0$ ensures that CG generates attractive gravity in the Newtonian limit~\cite{mannheim_2007}.
Taking the variation of~\eqref{eq:Weyl_action} with respect to the metric, we obtain
\begin{equation}\label{eq:variation}
\frac{-2}{\sqrt{-g}}\, \frac{\delta I_{\rm W}}{\delta g_{\mu \nu}} = 4 \, \alpha_g \, W^{\mu \nu},
\end{equation}
where the Bach tensor~\cite{mannheimgammavalue}:
\begin{equation}\label{eq:bach_tensor}
    W^{\mu\nu} = 2 {C^{\mu\lambda\nu\kappa}}_{;\lambda;\kappa} - {C^{\mu\lambda\nu\kappa}} \,R_{\lambda\kappa},
\end{equation}
is the CG analogue of the Einstein curvature tensor $G^{\mu\nu}$. From this, CG's Bach field equations~\cite{mannheimgammavalue} can be written:
\begin{equation}\label{eq:bach}
   4\,\alpha_g\, W_{\mu\nu} = T_{\mu\nu}.
\end{equation}
Importantly, because of the conformal invariance associated with the theory, both $W_{\mu\nu}$ and $T_{\mu\nu}$ must be traceless. This conformal invariance enforces gravity to be fourth-order, while this exact symmetry makes the SU(3)$\times$SU(2)$\times$U(1) theory of particle physics second-order~\cite{mannheimgammavalue}. Due to this, as well as its renormalisability~\cite{grumiller2014} and lack of negative-norm ghosts despite being a higher-order theory~\cite{bender2008_1, bender2008_2, bender2008_3}, CG has also been considered a viable candidate for a description of quantum gravity and as an alternative contender to string theory~\cite{mannheim2009}. 

Notwithstanding the advantages of CG listed above, several problems are currently unsolved in the theory; in astrophysics, orbit decays of binary pulsars have not been solved in CG~\cite{makingthecase}, 
and possible issues with gravitational lensing~\cite{edery1998, sophielensing} and rotation curves have been highlighted~\cite{KeithRotation, hobson_2021, hobson2022}. 
For conformal cosmology, big bang nucleosynthesis is a challenge: CG is able to produce enough helium, but has yet not been able to resolve the primordial deuterium problem~\cite{knox1993}. 
Moreover, CG may suffer from a similar finetuning problem to the $\Lambda\mathrm{CDM}$ model of GR, and it has been shown that the standard second-order theory provides a better fit to observations of high-redshift GRB and quasar standard candles~\cite{roberts2017}.

Because of the higher-order nature of the theory, CG admits exotic black hole solutions and spacetime structures that do not occur for the vacuum and electrovacuum metrics of GR, as seen in previous studies of the spacetime structures of the CG analogues~\cite{mannheim1989, mannheim1991} to GR's Schwarzschild~\cite{Schwarzschild} and rotating GR Kerr~\cite{Kerr1963} metrics~\cite{turner2020, yulo2025}. 
In the present work, we extend this analysis to CG's analogue of the GR Reissner-Nordstr\"{o}m metric~\cite{mannheim1991}. 

\subsection{CG's charged black hole metric}

The line element for static spherically symmetric nonrotating (electro)vacuum metrics in both GR and CG can be expressed as
\begin{equation}\label{eq:ds^2}
    \mathrm{d} s^2 = -B(r) \ \mathrm{d} t^2+\dfrac{\mathrm{d} r^2}{B(r)}+r^2 ( \mathrm{d}\theta^2 + \sin^2\theta \ \mathrm{d}\phi^2) ,
\end{equation} 
where $t$ and $r$ are the temporal and radial coordinates, and $\theta$ and $\phi$ are the polar and azimuthal angles respectively. For such metrics, the ``blackening factor'' $B(r)$ defines the causal structure of the spacetime. 
We may gain some intuition about whether terms in the blackening factor $B(r)$ act \textit{attractively} or \textit{repulsively} by considering the relation
\begin{equation}\label{eq:Phi(r)}
    B(r) = 1 + 2 \,\Phi(r),
\end{equation}
where $\Phi(r)$ may be interpreted as acting similarly to a Newtonian gravitational potential.

In GR, we have the following blackening factor for the uncharged case, referred to as the GR Schwarzschild (GRS) metric: 
\begin{equation}\label{eq:GRS}
    B_{\mathrm{GRS}}(r)=1-\dfrac{2\, \beta}{r}.
\end{equation}
Here, $\beta = G\,M/c^2$ is the gravitational radius, with $M$ denoting the central mass.

In the charged case, the GR Reissner-Nordstr\"{o}m metric (GRRN), the blackening factor is
\begin{equation}\label{eq:GRRN} 
    B_{\mathrm{GRRN}}(r)=1-\dfrac{2\,\beta}{r}+\dfrac{r_Q^2}{r^2} ,
\end{equation} 
where $r_Q^2=(Q^2 \,G/4\,\uppi\,\epsilon_0\,c^4)$ 
for charge $Q$. 
Setting $Q=0$ recovers the GRS metric. 
The metrics of~\eqref{eq:GRS} and~\eqref{eq:GRRN} have different dependences on $r$; for a nonzero charge we always obtain a timelike singularity $B(r=0)\rightarrow +\infty$ for the GRRN metric. In contrast, for the GRS metric, $B(r=0)\rightarrow-\infty$, and we always have a spacelike singularity as long as we restrict ourselves to the positive-mass solutions. 

In CG, the 4th-order field equations~\eqref{eq:bach} give a 4th-order Poisson equation~\cite{mannheimgammavalue, Brihaye_2009}:
\begin{equation}\label{eq:4_order}
\frac{3}{B}\,\left(W^0_0 - W^r_r \right) = \frac{1}{r}\left( r \, B \right)'''' =
\frac{3}{4\,\alpha_g\, B}\, \left( T^0_0 - T^r_r \right) \ ,
\end{equation}
where $'$ denotes ${\rm d}/{\rm d}r$. 
Since
$
(r^{n+1})'''' = (n+1)\,n\,(n-1)\,(n-2)\,r^{n-3}\ ,
$
the homogeneous 4th-order Poisson equation~\eqref{eq:4_order} with $T^0_0=T^r_r$ gives solutions of the form
\begin{equation}\label{eq:CG_nonrot_metric}
    B(r)=w+\dfrac{u}{r}+v\,r-k\,r^2 
    \ .
\end{equation}
Here the 4 integration constants give rise to the constant $w$, Newtonian $u/r$, linear $v\,r$, and quadratic $k\,r^2$ potentials.
The 4 parameters in~\eqref{eq:CG_nonrot_metric} must also satisfy a 3rd-order constraint 
arising from the $rr$-component of the Bach equations~\eqref{eq:bach}~\cite{KeithRotation, Brihaye_2009}: 
\begin{equation}\label{eq:W_rr}
\begin{aligned}
   W^r_r =&\dfrac{1-B^2}{3\,r^4}+\dfrac{2\,B\,B'}{3\,r^3}-\dfrac{B\,B''+(B')^2}{3\,r^2}
    +\dfrac{B'B'' - BB'''}{3\,r}+\dfrac{2\,B'B'''-(B'')^2}{12} = \frac{ T^r_r }{ 4 \, \alpha_g}. 
\end{aligned}
\end{equation}
Substituting~\eqref{eq:CG_nonrot_metric} into
\eqref{eq:W_rr} gives
\begin{equation}\label{eq:CG_third_order_uncharged}
    w^2 = 1+3\,u\,v - \frac{3\,r^4}{4\,\alpha_g}\, T^r_r \ ,
\end{equation}
removing one degree of freedom in the choice of CG parameters. 

The CG Schwarzschild (CGS) metric is the source-free vacuum
solution with $T^\mu_\nu=0$, taking the form 
\cite{mannheim1989}
\begin{equation}\label{eq:CGS_params}
    w = 1-3\,\beta\,\gamma\ ,\ \quad
    u = -\beta\,(2-3\,\beta\,\gamma)\ ,\quad
    v = \gamma \ , \quad
    k = \kappa \ ,
\end{equation}
to give the 3-parameter CGS metric
\begin{equation}\label{eq:MK_metric}
    B_{\mathrm{CGS}}(r)=(1-3\,\beta\,\gamma) -\dfrac{\beta\,(2-3\,\beta\,\gamma)}{r}+\gamma\, r-\kappa\, r^2 .
\end{equation} 
Equation~\eqref{eq:MK_metric} encodes not only the constant and $1/r$ terms seen in the GRS metric~\eqref{eq:GRS}, but also a $\kappa\,r^2$ term seen in GR metrics with de Sitter (dS, $\kappa>0$) and anti-de Sitter (AdS, $\kappa<0$) cosmological backgrounds~\cite{mannheim1989, riegert1984}. Thus, by fiducially setting $\kappa=\Lambda/3$, the CGS metric may be understood to represent a GRS(A)dS metric modified by the contribution of the linear $\gamma\, r$ and constant $-3\,\beta\,\gamma$ terms. 
The linear $\gamma \,r$ term allows CG to fit a wide variety of observed galaxy rotation curves~\cite{mannheim_2022}; such linear terms in the blackening factor are also seen in Weyl Connection Gravity and their associated BH solutions~\cite{gomes2019, lima2024}.
For $\gamma=0$, the mass-like parameter $\beta$ reduces to the Newtonian mass. 
It is then clear that CGS~\eqref{eq:MK_metric} reduces to GRS(A)dS for $\gamma = 0$, and further to GRS~\eqref{eq:GRS} when both $\gamma = \kappa = 0$~\cite{mannheim1989}. 

For a non-rotating charged black hole, the
CG Reissner-Nordstr\"{o}m (CGRN) metric
is obtained by solving the Weyl-Maxwell system of equations with an appropriate vector potential~\cite{mannheim1991}: 
\begin{equation}\label{eq:vec_pot}
    A_\mu = (Q/r, 0,0,-P\cos\theta)\ ,
\end{equation}
where $Q$ and $P$ are the electric and magnetic charges, respectively. 
Due to its tracelessness, $T^\mu_\nu$ has only one independent component, which is given by
\begin{equation}\label{eq:T_rr}
    T^0_0=T^r_r=-\dfrac{Q^2 + P^2}{2\,r^4}\ .
\end{equation}
Using this, the 3rd-order constraint of~\eqref{eq:W_rr} can be solved to give~\cite{mannheim1991}:
\begin{equation}\label{eq:CG_third_order_charged}
    w^2 = 1+3\,u\,v
    - D_g^2\ ,
\end{equation}
 where here we define a dimensionless and positive ``dyonic'' parameter
 \begin{equation}
     D_g^2\equiv\dfrac{-3\,(Q^2+P^2)}{8\,\alpha_g},
 \end{equation}
 incorporating both charges and the negative gravitational coupling constant $\alpha_g$.
There is a two-dimensional manifold of options to satisfy~\eqref{eq:CG_third_order_charged} 
by assigning $D_g^2$ to mixtures of $w$, $u$, and $v$: 
\begin{equation} \label{eq:mess} 
w^2 = \left( 1 - p\, D_g^2 \right)
+ 3 \, \left( u - q \, \frac{ D_g^2} { 3 \, v } \right)\, v
+ 3 \, u \, \left( v + \left( p + q - 1 \right) \, \frac{D_g^2}{3 \, u} \right) \ .
\end{equation}
This degeneracy can be remedied by matching an interior CGRN solution to the exterior solution~\cite{mannheim2025_perscomm}.
Rather than tackling the interior solution,
or coping with the complexity of~\eqref{eq:mess},
we adopt the choices of Mannheim and Kazanas in~\cite{mannheim1991}, namely $(p,q)=(0,1)$ if $v\neq0$ and $(p,q)=(1,0)$ if $v=0$. 
 
Following~\cite{mannheim1991}, when $\gamma \neq 0$ the dyonic charge term $D_g^2$ modifies the mass parameter $u$:
\begin{equation}\label{eq:CGRN_params}
w = 1-3\,\beta\,\gamma 
\ , \quad
   u = -\beta\,(2-3\,\beta\,\gamma)
   + \frac{ D_g^2 }{3 \, \gamma}
\ , \quad
    v = \gamma
\ , \quad
    k = \kappa
\ .
\end{equation} 
The blackening factor for the $\gamma \neq 0$ CGRN metric is then
\begin{equation}\label{eq:CGRN_metric_1}
    B^{\gamma \neq0}_{\mathrm{CGRN}}(r)=\left( 1-3\,\beta\,\gamma \right)
    -\dfrac{1}{r}\,\left(\beta\,(2-3\,\beta\,\gamma) -\dfrac{D_g^2}{3\,\gamma}\right)+\gamma \,r-\kappa\, r^2. 
\end{equation}
This reduces to $B_{\mathrm{CGS}}(r)$~\eqref{eq:MK_metric} in the charge-free limit $D_g = 0$. 

Again following~\cite{mannheim1991}, for $\gamma=0$ the four parameters of~\eqref{eq:CG_nonrot_metric} are
\begin{equation}\label{eq:CGRN_params_g0} 
    w = 
     w_0 
\ , \quad
    u = -2\,\beta
\ , \quad
    v = 0
\ , \quad  
    k = \kappa
\ ,
\end{equation}
where we define
\begin{equation}\label{eq:w0}
    w_{0}^2\equiv1-\,D_g^2 \ .
\end{equation}

The blackening factor in this case is: 
\begin{equation}\label{eq:CGRN_metric_2}
    B_{\mathrm{CGRN}}^{\gamma = 0}(r)=w_0-\dfrac{2\,\beta}{r}-\kappa \,r^2 \ .
\end{equation}
To ensure that~\eqref{eq:CGRN_metric_2} is asymptotically Minkowski when $\kappa$ and $D_g^2$ are set to zero, we define $w_0$ as the positive root of~\eqref{eq:w0}.~\eqref{eq:CGRN_metric_2} acts like a GRS(A)dS metric with its constant term reduced from 1 to $w_0$.
It requires an upper limit $D_g^2\leq1$ to avoid an imaginary term in $B(r)$.
Note that~\eqref{eq:CGRN_metric_1} does not reduce to~\eqref{eq:CGRN_metric_2} in the limit $\gamma\rightarrow0$. 
We therefore need to treat these two metrics separately. 
Henceforth, when we refer to a general $B(r)$, we refer to either~\eqref{eq:CGRN_metric_1} or~\eqref{eq:CGRN_metric_2}. 

The GRRN metric~\eqref{eq:GRRN} has a $1/r^2$ term that is absent in the CGRN metric.  
Thus even with $\gamma=0$ the CGRN metric does \textit{not} reduce to the GRRN(A)dS metric,
and further setting $\kappa = 0$ does not recover GRRN. 
This is in direct contrast to the CGS~\eqref{eq:MK_metric} and CG Kerr~\cite{mannheim1991} metrics which \textit{do} reduce to GRS(A)dS and GR\,Kerr\,(A)dS respectively for $\gamma=0$. 
These further reduce to GRS and GR Kerr respectively when both $\gamma$ and $\kappa$ vanish~\cite{yulo2025,
VarieschiFlyby}. 
As these reductions to GR hold for uncharged metrics, they point to how handling charge in CG may be fundamentally different from how it is treated in GR~\footnote{We note the existence of papers~\cite{payandeh2012, fathi2020} which claim that a charged black hole metric in CG involves a charge term going as $1/r^2$. This claim arises from the assumption that the second-order Poisson equation is viable in the present fourth-order theory, which as stated instead requires a fourth-order Poisson equation~\eqref{eq:4_order}.}.

As highlighted by Mannheim and Kazanas, Reissner and Nordstr\"{o}m intended to obtain a theory unifying gravitation and electromagnetism. Therefore, Mannheim and Kazanas claim that while the $1/r^2$ term in GRRN undesirably gives different behaviours for mass and charge, no such problem is encountered in its conformal counterpart when $\gamma\neq0$~\cite{mannheim1991}. 

Surprisingly, perhaps due to the interests of relativists skewing towards astrophysical applications of CG, wherein it is assumed that the central supermassive black holes at the centres of galaxies are uncharged, the CGRN metric has not garnered much attention. Hence, with the present work, we present a comprehensive analysis of the various exotic spacetimes obtainable in the CGRN solution, which describes nonrotating electrovacuum dyonic metrics. We first briefly discuss the terminologies we apply to discuss the spacetime features of nonrotating metrics and black holes (section~\ref{sec:terminologies}), and then present expressions for relevant curvature scalars of the CGRN metric (section~\ref{sec:singularities}). In section~\ref{sec:null_geodesics}, we further present equations of motion for null particles. Then, we discuss horizon structures of the metric (section~\ref{sec:horizons}), and present a full review of the different spacetime configurations in section~\ref{sec:spacetime_domains}. We conclude and provide further perspectives in section~\ref{sec:concl}. 

Throughout this work, we adopt dimensionless parameters $r/\beta$, $\beta\,\gamma$ and $\beta^2 \,\kappa$. Additionally, as $\beta<0$ is often associated with exotic matter that generates negative energy density or pressure~\cite{turner2020}, we restrict ourselves to positive mass $\beta>0$ in the present work, as well as attractive gravity in the Newtonian limit ($\alpha_g<0$)~\cite{mannheim_2007}. Hence, the sign of the dimensionless parameter $\beta\gamma$ is governed by the sign of $\gamma$ throughout the manuscript.

\section{Terminology for classifying static spacetimes}\label{sec:terminologies}

\subsection{Classifications of spacetime regions}

We refer throughout this work to \textit{timelike} (T) and \textit{spacelike} (S) regions of spacetime. 
Timelike ($\mathrm{T}$) regions are where $g_{rr} = B(r) > 0$. 
In T\,regions, $\mathrm{d}t$ is a timelike interval, and hence $t$ is a timelike coordinate. The future world lines of particles, $\mathrm{d}t>0$, are permitted to progress to increasing or to decreasing $r$. 

Spacelike (S) regions 
are where $g_{rr} = B(r) < 0$.
Here $\mathrm{d}t$ is spacelike and $\mathrm{d}r$ is timelike. In this sense the $t$ and $r$ coordinates switch roles. S regions come in two types ($\mathrm{S}^-$ and $\mathrm{S}^+$).
In $\mathrm{S}^{-}$ regions, $\mathrm{d}r<0$ is in
the future light cone and particles must progress towards decreasing $r$.
A familiar example is the $\mathrm{S}^{-}$ region inside the event horizon of a GRS black hole.
In the spacelike regions that we denote as $\mathrm{S}^{+}$, 
$\mathrm{d}r>0$ is in the future light cone,
and particles must progress toward increasing $r$.
An $\mathrm{S}^{+}$ region is found exterior to cosmological horizons, such as those present in the GRSdS metric~\cite{yulo2025}.

\subsection{Types of horizons}\label{subsec:horizon_types}

For the sake of conciseness throughout the rest of the present work, we abbreviate the usual horizon types as follows: 
\begin{itemize}
    \item Event horizon $\mathrm{H}_\mathrm{E}$: For increasing $r$, an event horizon separates an interior S$^-$ region from an exterior T region ($\mathrm{S}^-\rightarrow \mathrm{H}_\mathrm{E} \rightarrow \mathrm{T}$). The black hole event horizons of GRS spacetimes are perhaps the most familiar example. We then say that a spacetime has a black hole if $\mathrm{H}_{\mathrm{E}}$ and its corresponding $\mathrm{S}^-$ region are present. 
    \item Cauchy horizon $\mathrm{H}_\mathrm{C}$: For increasing $r$, a Cauchy horizon separates an interior T region from an exterior S$^-$ region ($\mathrm{T} \rightarrow \mathrm{H}_\mathrm{C} \rightarrow \mathrm{S}^-$). Technically, Cauchy horizons demarcate where Cauchy information suffices to determine the causal past of an event. They are thus often associated with the innermost horizons of rotating or charged black holes in GR. In these cases, the Cauchy horizons separate a $\mathrm{T}$ region from the $\mathrm{S}^-$ region within the black hole's event horizon $\mathrm{H}_\mathrm{E}$. Since the $\mathrm{T}$ region contains the singularity in such cases, the casual past of events within the $\mathrm{T}$ region can no longer be determined solely by Cauchy information. However, in the CGS and CGRN metrics, we will encounter spacetimes wherein $\mathrm{H}_\mathrm{C}$ does not surround the central singularity, Therefore, for our purposes, we crudely use this designation of Cauchy horizon $\mathrm{H}_\mathrm{C}$ to simply define horizons demarcating the aforementioned progression ($\mathrm{T} \rightarrow \mathrm{H}_\mathrm{C} \rightarrow \mathrm{S}^-$). 
    \item Cosmological horizon $\mathrm{H}_\Lambda$: For increasing $r$, a cosmological horizon separates an interior T region from an exterior S$^+$ region ($\mathrm{T}\rightarrow \mathrm{H}_{\Lambda} \rightarrow \mathrm{S}^+$). Like a Cauchy horizon $\mathrm{H}_{\mathrm{C}}$, we can think of cosmological horizons as being generated by an \textit{effective repulsion}. While Cauchy horizons carve out a $\mathrm{T}$ region within an $\mathrm{S}^-$ region, cosmological horizons bound $\mathrm{S}^+$ regions from $\mathrm{T}$ regions. Notably, cosmological horizons are usually the result of a positive $\beta^2\kappa$, as in the GRSdS metric. However, in the static nonrotating spacetimes of CG~\eqref{eq:CG_nonrot_metric}, one may still obtain a cosmological horizon even if $\beta^2 \kappa=0$, if $\beta\,\gamma<0$. This is because with the vanishing of the quadratic $-\kappa \,r^2$ term in $B(r)$, the linear $\gamma\, r$ dominates the large $r$ behavior.

\end{itemize}
We additionally introduce the notation used in~\cite{turner2020} of \textit{in} (I) and \textit{out} (O) horizons for ease of comparison to their mapping of the CGS spacetimes. In this notation, an event horizon $\mathrm{H}_\mathrm{E}$ is classified as I to indicate an S region ($B<0)$ interior to a T region ($B>0$),
and thus $B'(r)>0$ at the horizon. Cosmological and Cauchy horizons are classified as O because they separate an S region outside a T region ($B'(r)<0$). These classifications become most relevant in section~\ref{sec:spacetime_domains}. 

Furthermore, we identify 4 types of horizon associated with extremal limits of the CGRN metric: the \textit{extremal charge} $\mathrm{H}_\mathrm{CE}$ horizon (where $\mathrm{H} _\mathrm{C}$ inside $\mathrm{H}_\mathrm{E}$ merge),
the \textit{nested limit} horizon $\mathrm{H}_\mathrm{EC}$ ($\mathrm{H}_\mathrm{E}$ inside $\mathrm{H}_\mathrm{C}$ merge), the \textit{Nariai} $\mathrm{H}_{\mathrm{E}\Lambda}$ horizon
($\mathrm{H}_\mathrm{E}$ inside $\mathrm{H}_\mathrm{\Lambda}$ merge), and the \textit{extremal triple limit} horizon $\mathrm{H}_\mathrm{TL}$
(all three horizons of the metric coincide).

The extremal charge horizon $\mathrm{H}_\mathrm{CE}$ forms when an interior Cauchy horizon $\mathrm{H}_\mathrm{C}$ 
(at radius $r_\mathrm{H_C}$) and an exterior event horizon $\mathrm{H}_\mathrm{E}$ (at
radius $r_\mathrm{H_E}$) come together and merge
$(r_{\mathrm{H_C}} < r_{\mathrm{H_E}})$. 
In GRRN spacetimes, this limit is reached when the magnitude of the charge exactly balances that of the mass as $\beta = r_Q$; exceeding this limit violates the weak cosmic censorship hypothesis. In CGRN, we shall also see an additional dependence of such extremal limits on $\gamma$ and $\kappa$, but we retain the terminology of \textit{extremal charge} from GRRN for simplicity. 

The nested limit horizon $\mathrm{H}_\mathrm{EC}$ similarly forms from the coalescence of an event and Cauchy horizon, like the aforementioned extremal charge limit $\mathrm{H}_\mathrm{CE}$. However, in this case the event horizon is interior to the Cauchy horizon $(r_{\mathrm{H_E}} < r_{\mathrm{H_C}})$. While not often discussed, we encounter this in CGS and CGRN spacetimes, as will be shown later in section~\ref{sec:spacetime_domains}. 

In GR black hole spacetimes with de Sitter backgrounds, the near horizon geometry at the limit where the event $\mathrm{H}_\mathrm{E}$ and cosmological $\mathrm{H}_\Lambda$ horizons coalesce may be shown to reduce to the Nariai solution~\cite{NariaiOriginal, NariaiHolo}. We, however, do not show that such an analogue to the Nariai solution is recovered at this limit in CG. We denote it here as $\mathrm{H}_{\mathrm{E}\Lambda}$, referring to the coalescence of $\mathrm{H}_\mathrm{E}$ and $\mathrm{H}_\Lambda$. 

The last extremal case we encounter, the triple limit, occurs where three horizons coincide. Such a limit is found in some GR spacetimes with dS backgrounds, where the Cauchy $\mathrm{H}_\mathrm{C}$, event $\mathrm{H}_\mathrm{E}$, and cosmological $\mathrm{H}_\Lambda$ horizons merge in what is known as the \textit{ultracold limit}~\cite{castro2023}. However, since we shall see cases in CGRN where we have no cosmological horizon $\mathrm{H}_\Lambda$ but instead have two event horizons $\mathrm{H}_\mathrm{E}$ and one Cauchy horizon $\mathrm{H}_\mathrm{C}$ coinciding $(r_{\mathrm{H_E, interior}} < r_{\mathrm{H_C}}<r_{\mathrm{H_E, exterior}})$, we do not use the term ultracold limit here.

\section{Small \texorpdfstring{$r$}{r} behaviour and curvature invariants}\label{sec:singularities}

\subsection{Role of charge and small \texorpdfstring{$r$}{r} behaviour}\label{subsec:role of charge}

Charge is always repulsive in GRRN, due to the positive sign of $r_{Q}^2/r^2$ in  $B_{\mathrm{GRRN}}(r)$~\eqref{eq:GRRN}, as can be seen in~\eqref{eq:Phi(r)}. This leads to the singularity at $r = 0$ always being timelike, as $B_{\mathrm{GRRN}}(r=0)\rightarrow+\infty$. While the mass $\beta$ acts attractively in the $-(2\,\beta/r)$ term of $B_{\mathrm{GRRN}}(r)$~\eqref{eq:GRRN}, the repulsive charge term $+(r_Q^2/r^2)$ dominates the behaviour of $B_\mathrm{GRRN}(r)$ at small $r$. Thus, below the extremal limit, this repulsion generates a Cauchy horizon $\mathrm{H}_{\mathrm{C}}$ which carves out a $\mathrm{T}$ region interior to the $\mathrm{S}^-$ region of the black hole. Past the extremal limit of GRRN, all of spacetime is timelike.

Things are more complicated in the $\gamma \neq 0$ case of CGRN. 
The dyonic charge does not generate a separate $1/r^2$ term in $B^{\gamma \neq 0}_{\mathrm{CGRN}}(r)$~\eqref{eq:CGRN_metric_1}, but instead modifies the $1/r$ term. The coefficient of the $1/r$ term corresponding to the charge appears as $+(D_g^2/3\,\gamma)$, which can change sign with the sign of $\beta \gamma$. Thus, unlike in GRRN, the central singularity is not always timelike. 

\begin{figure}
    \centering
    \includegraphics[width=0.6\linewidth]{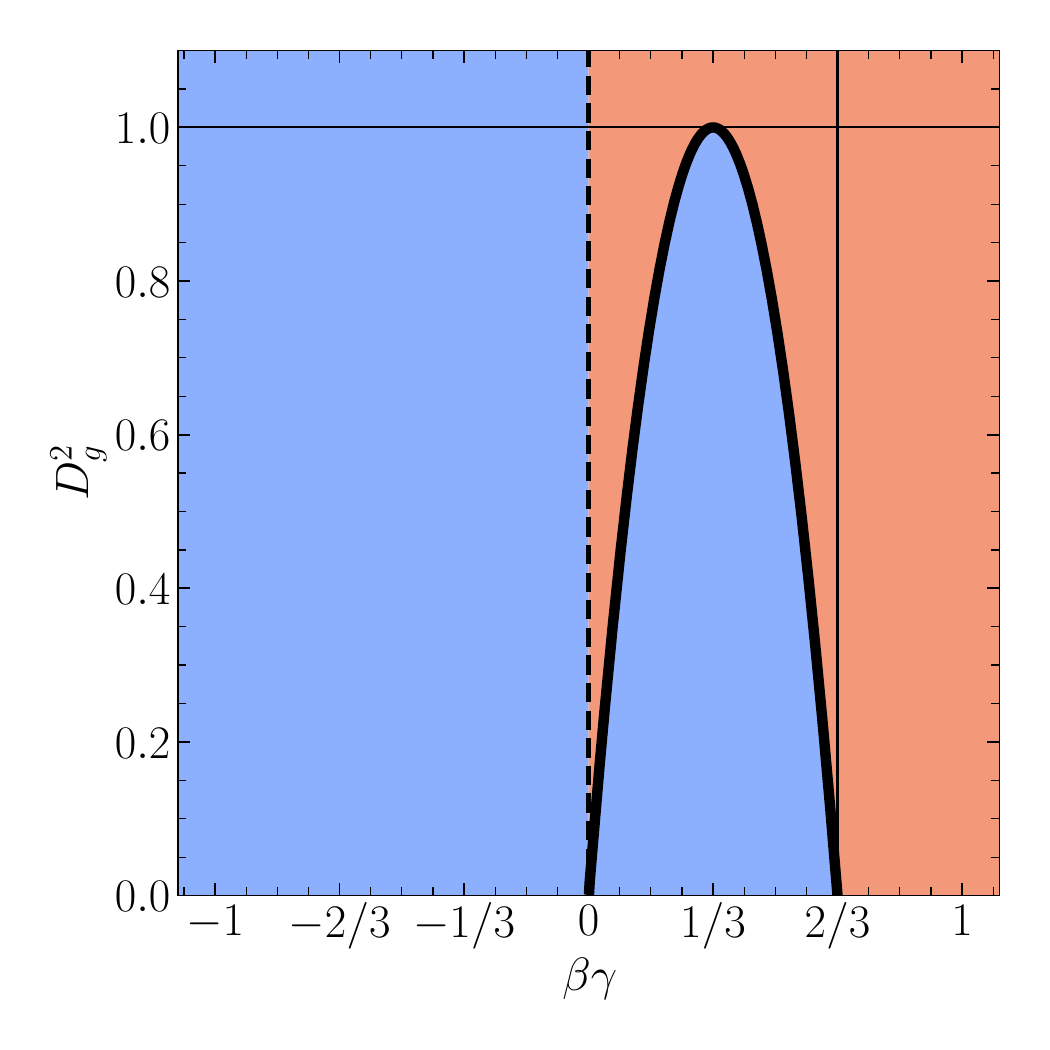}
    \caption{Diagram showing the nature of the singularity in the CGRN metric, for given combinations of $D_g^2$ and $\beta\gamma$. Blue regions correspond to spacelike singularities ($B(0)\rightarrow-\infty$), red regions correspond to timelike singularities ($B(0)\rightarrow+\infty$), and the thick black contour that demarcates these regions corresponds to spacetimes with no singularities. $B^{\gamma\neq0}_\mathrm{CGRN}(r)$~\eqref{eq:CGRN_metric_1} is ill-defined at $\beta\gamma=0$, so we notate this discontinuity by a dashed line.} 
    \label{fig:small_r_singul}
\end{figure}

When $\beta\,\gamma<0$, both $-(2-3\,\beta\,\gamma)$ and $+(D_g^2/3\,\beta\,\gamma)$ are negative, and thus the central singularity is spacelike. If $\beta\gamma>2/3$, $-(2-3\,\beta\,\gamma)>0$ and $+(D_g^2/3\,\beta\,\gamma)>0$, so the central singularity is timelike with $B(r=0)\rightarrow+\infty$. For $0<\beta\,\gamma\leq2/3$, it is the relative magnitudes of $\beta\gamma$ and $D_g^2$ which determines whether the singularity is spacelike or timelike. This is shown in figure~\ref{fig:small_r_singul}, where it is evident that the competition between the charge and $\beta\gamma$ is important in dictating the nature of the central singularity. The figure also highlights the critical value of $D_g^2=1$ at the tip of the parabolic border, above which all singularities are timelike if $\beta\gamma>0$.

On the other hand, the nature of the singularity in the $\gamma=0$ metric~\eqref{eq:CGRN_metric_2} 
is more straightforward as this is dictated solely by the sign of $\beta$. 
For $\beta>0$, the singularity is spacelike.
Meanwhile, timelike singularities are exclusively found when $\beta<0$, but we do not discuss such negative-mass solutions here. 

\subsection{Curvature scalars}

To find physical singularities present in the spacetimes, we calculate curvature scalars of the metric using the newly developed \texttt{OGRePy}\footnote{\href{https://github.com/bshoshany/OGRePy/tree/master}{https://github.com/bshoshany/OGRePy/tree/master}}, a Python implementation of the Mathematica package used for tensor calculus~\cite{shoshany2025}.

The first of these curvature scalars that we discuss is the Ricci scalar $R={R^\mu}_\mu$, obtained from contracting the Ricci tensor $R_{\mu\nu}$. For a general blackening factor of the form~\eqref{eq:CG_nonrot_metric}, which we obtain for nonrotating metrics in CG, $R$ is generally written
\begin{equation}\label{eq:CG_Ricci_scalar}
    R = \dfrac{2\,(1-w)}{r^2} - \dfrac{6\,v}{r} + 12\,k.
\end{equation}
Inserting the parameters presented in~\eqref{eq:CGRN_params} for the $\gamma\neq0$ metric into~\eqref{eq:CG_Ricci_scalar}, we then obtain 
\begin{equation}\label{eq:CGRN_Ricci}
    R^{\gamma\neq0}=\dfrac{6\,\beta\,\gamma}{r^2}-\dfrac{6\,\gamma}{r}+12\,\kappa,
\end{equation} 
which is exactly equal to $R$ for the uncharged CGS metric~\cite{turner2020}. This is to be expected; $R$ is also the same in both GRS and GRRN metrics. Much like in the CGS metric, all metrics with $\gamma\neq0$ possess physical singularities, due to the divergence of $R$ at $r=0$~\cite{turner2020}. 

Next, taking the Kretschmann scalar $K=R^{\mu\nu\rho\sigma}R_{\mu\nu\rho\sigma}$ with $R_{\mu\nu\rho\sigma}$ denoting the Riemann tensor, we have 
\begin{equation}\label{eq:CG_Kretschmann_scalar} 
    K=24k^2 - \dfrac{24kv}{r} -\dfrac{8(kw - k - v^2)}{r^2} + \dfrac{8v(w-1)}{r^3} + \dfrac{4(w-1)^2}{r^4} + \dfrac{8u(w-1)}{r^5} + \dfrac{12u^2}{r^6}. 
\end{equation} 
Evaluating this gives 
\begin{equation}\label{eq:CGRN_Kretschmann} 
\begin{aligned} 
    K^{\gamma\neq0}= 24\,\kappa^2 &- \dfrac{24\,\gamma\,\kappa}{r} + \dfrac{24\,\beta\,\gamma\,\kappa + 8\,\gamma^2}{r^2} - \dfrac{24\,\beta\,\gamma^2}{r^3} + \dfrac{36\,\beta^2\,\gamma^2}{r^4} \\&+ \dfrac{24\,\beta^2\,\gamma\,(2-3\,\beta\,\gamma) - 8\,\beta \,D_g^2}{r^5} + 
    \dfrac{1}{r^6}\, \left(-\beta\,(2-3\,\beta\,\gamma) + \dfrac{D_g^2}{3\,\gamma}\right)^2, 
\end{aligned} 
\end{equation} 
affirming that the $\gamma\neq0$ metric of~\eqref{eq:CGRN_metric_1} is always singular at $r=0$.

From~\eqref{eq:CGRN_Ricci} and~\eqref{eq:CGRN_Kretschmann} we see that the $\gamma\neq0$ metric given in~\eqref{eq:CGRN_metric_1} always possesses a true curvature singularity in the standard Schwarzschild frame where $g_{tt}=-1/g_{rr}$. 

On the other hand, evaluating $R$ for the $\gamma=0$ metric given in~\eqref{eq:CGRN_metric_2}, we find 
\begin{equation}
    R^{\gamma=0}=\dfrac{2\left(1 - w_0\right)}{r^2}+12\kappa,
\end{equation}
and the Kretschmann scalar $K$ is
\begin{equation}
    K^{\gamma=0}=24\,\kappa^2 - \dfrac{8\,\kappa\,\left(w_0-1\right)}{r^2}+\dfrac{4\,\left(w_0-1\right)^2}{r^4}-\dfrac{16\,\beta\,\left(w_0-1\right)}{r^5}+\dfrac{48\,\beta^2}{r^6}, 
\end{equation}
where $w_0$ is as defined in~\eqref{eq:w0}. From these two curvature scalars, the singularity is also unavoidable unless charge and mass simultaneously vanish as $D_g = 0 \rightarrow w_0 = 1$ and $\beta = 0$.

Note that these curvature scalars are calculated in the Schwarzschild frame where $g_{tt}=-1/g_{rr}$; as~\cite{turner2020} remarks, a conformal Weyl transformation to another frame may be able to remove curvature singularities from a metric in a conformal theory of gravity.

\section{Null geodesics and photon spheres}\label{sec:null_geodesics}

Trajectories of null particles in CGRN spacetimes can be obtained from the null geodesic equation and the Killing vectors representing conserved quantities. The Killing vectors $\partial_t$ and $\partial_\phi$ give rise to the conserved quantities $E$ and $L$ respectively. Technically, as CGRN spacetimes are not generally asymptotically flat, $E$ and $L$ do not represent the energy and angular momentum directly. 

\begin{figure}
    \centering
    \begin{subfigure}[b]{0.99\textwidth} 
        \includegraphics[width=\textwidth]{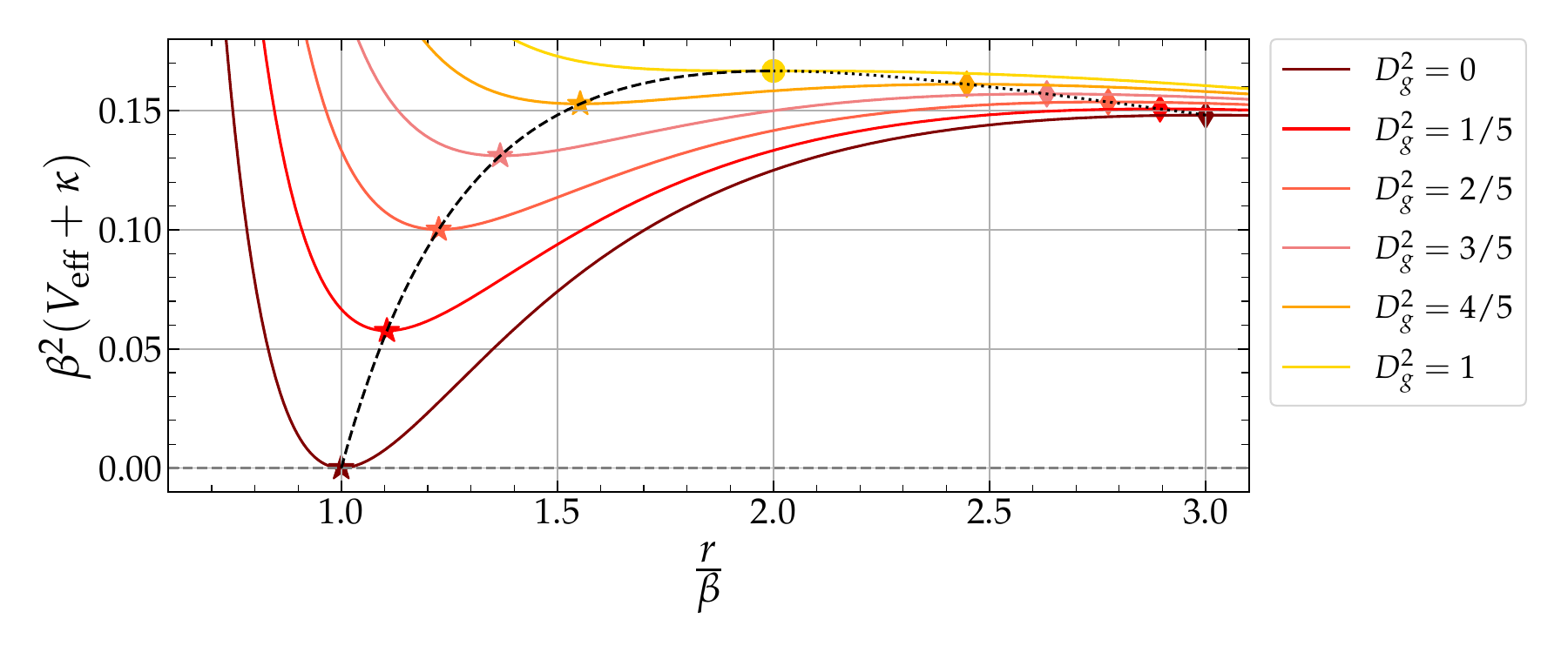} 
        \caption[]%
        {$\beta\gamma=1$, $\beta^2\kappa=0$.} 
        \label{subfig:V_eff_bg+}
    \end{subfigure}
    \hfill
    \begin{subfigure}[b]{0.99\textwidth}  
        \includegraphics[width=\textwidth]{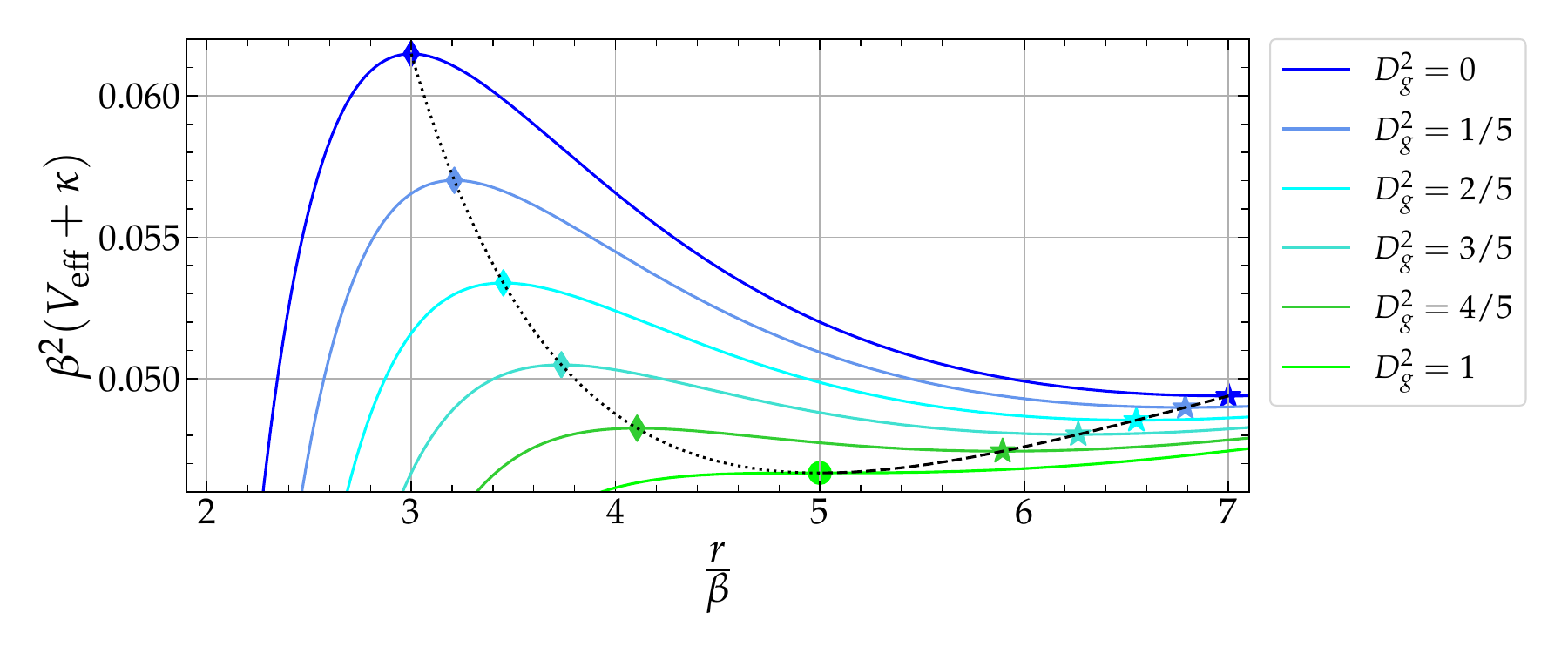}
        \caption[]%
        {$\beta\gamma=-0.5$, $\beta^2\kappa=-0.08$.} 
        \label{subfig:V_eff_bg-}
    \end{subfigure}
    \caption{The effective potential $V_\mathrm{eff}$~\eqref{eq:V_eff} for null particles in the $\gamma\neq0$ CGRN metric~\eqref{eq:CGRN_metric_1} against $r/\beta$, for different values of charge $D_g$. Stars and diamonds correspond to stable and unstable photon spheres $r_\mathrm{st}$ and $r_\mathrm{ust}$~\eqref{eq:photonspheres} respectively, and the circles denote the marginally stable \textit{saddle point} photon sphere $r_\mathrm{mst}$~\eqref{eq:saddle}. The black dashed and dotted lines show the locations of the stable and unstable photon spheres as charge is increased to its maximum value of $D_g^2=1$.} 
    \label{fig:V_eff} 
\end{figure}

As $E$ is conserved over time $t$ and $L$ is conserved over azimuth $\phi$, we obtain four equations of motion with respect to an affine parameter $\lambda$~\cite{stewart1989}: 
\begin{equation} 
\begin{aligned} 
    \dfrac{\mathrm{d}t}{\mathrm{d}\lambda} &= \dfrac{E}{B(r)},\\
    \left(\dfrac{\mathrm{d}r}{\mathrm{d}\lambda}\right)^2&= E^2 - \dfrac{r^2 B(r)}{2} \left(\dfrac{\mathrm{d}\theta}{\mathrm{d}\lambda}\right)^2-\dfrac{L^2B(r)}{r^2\sin^2\theta},\\
     \dfrac{\mathrm{d}}{\mathrm{d}\lambda}\left(r^2 \dfrac{\mathrm{d}\theta}{\mathrm{d}\lambda}\right)&= \dfrac{L^2 \cos\theta}{r^2\sin^3\theta},\\
    \dfrac{\mathrm{d}\phi}{\mathrm{d}\lambda} &= \dfrac{L}{r^2 \sin^2\theta}\ .
\end{aligned}
\end{equation}
Considering just the equatorial plane and combining the $r$ and $\phi$ derivatives, we have 
\begin{equation}\label{eq:vel} 
    \left(\dfrac{\mathrm{d}r}{\mathrm{d}\phi}\right)^2 = \dfrac{r^4}{L^2}\left(\,E^2 - L^2\,V_\text{eff}(r)\,\right)\ ,
\end{equation}
where $V_\mathrm{eff}(r)=B(r)/r^2$ is the effective potential for null particles. For $\gamma\neq0$, we have 
\begin{equation}\label{eq:V_eff}
    V^{\gamma\neq0}_{\text{eff}}(r)=\frac{ 1-3\,\beta\,\gamma}{r^2}   
   -\frac{\beta\,(\,2-3\,\beta\,\gamma\,) -\dfrac{D_g^2}{3\gamma}}{r^3}+\frac{\gamma}{r}-\kappa \ ,
\end{equation}
while for the $\gamma=0$ metric we obtain 
\begin{equation}\label{eq:V_eff_gamma0}
    V^{\gamma=0}_{\text{eff}}(r)=\frac{w_0}{r^2}-\frac{2\,\beta}{r^3}-\kappa \ . 
\end{equation} 
Figure~\ref{fig:V_eff} shows some examples of~\eqref{eq:V_eff} for different values of $D_g^2$, $\beta\gamma$, and $\beta^2\kappa$. 

Photon spheres, which are circular null geodesics, occur at the extrema of $V_\mathrm{eff}$, where $\rmd V_{\text{eff}}/\rmd r = 0$~\cite{turner2020}. This condition is given for $\gamma \neq 0$ by

\begin{equation}\label{eq:dV}
\frac{\mathrm{d} V^{\gamma\neq0}_{\text{eff}}}{\mathrm{d} r} = -\frac{2\,(1-3\,\beta\,\gamma)}{r^3} + \frac{3}{r^4}\,\left(\beta\,(2-3\,\beta\,\gamma) -\dfrac{D_g^2}{3\gamma}\right) - \frac{\gamma}{r^2} = 0
\ ,
\end{equation}
while for $\gamma = 0$ we have
\begin{equation}\label{eq:dVgammazero}
\frac{\mathrm{d} V^{\gamma=0}_{\text{eff}}}{\mathrm{d} r} = -\frac{2\,w_0}{r^3}+ \frac{6\,\beta}{r^4} = 0
\ .
\end{equation}

Much like the CGS metric~\eqref{eq:MK_metric} which has stable and unstable photon spheres located at $r=3\beta-2/\gamma$ and $r=3\beta$ respectively~\cite{turner2020, kusano2025}, the CGRN metric also allows for stable and unstable photon orbits.


By solving for photon sphere radii in the $\gamma\neq0$ CGRN metric using~\eqref{eq:dV}, we find
\begin{equation}
    \begin{aligned}\label{eq: r photon}
    \dfrac{r}{\beta}
    &= \dfrac{3\,\beta\,\gamma-1\pm w_0}{\beta\,\gamma}\ ,
\end{aligned}
\end{equation}
which, as expected, reduces to the expression for the CGS metric photon sphere radii when the dyonic parameter $D_g^2=0$~\cite{turner2020}. From this, we see that
\begin{equation}\label{eq:photonspheres}
    \begin{aligned}
        \dfrac{r_\mathrm{st}}{\beta}&=\dfrac{3\,\beta\,\gamma-1-w_0}{\beta\,\gamma},\\
        \dfrac{r_\mathrm{ust}}{\beta}&=\dfrac{3\,\beta\,\gamma-1+w_0}{\beta\,\gamma},
    \end{aligned}
\end{equation}
with $r_\mathrm{st}$ and $r_\mathrm{ust}$ corresponding to stable and unstable photon sphere radii respectively. 
While it initially seems from~\eqref{eq:photonspheres} as though the stable photon sphere is interior to the unstable one, note that this is not necessarily the case, 
depending on the sign of $\beta\,\gamma$.

Recall that $w_0^2 =1 - D_g^2$ implies an upper limit $D_g^2 \leq 1$. As $D_g^2\rightarrow1$, $w_0\rightarrow0$ and the two photon spheres~\eqref{eq: r photon} merge to form a single photon sphere at
\begin{equation}\label{eq:saddle}
    \frac{r_\mathrm{mst}}{\beta}=\frac{3\,\beta\, \gamma - 1}{\beta\,\gamma}
    \ .
\end{equation}
As the second derivative of $V^{\gamma\neq0}_{\text{eff}}(r)$ also vanishes at $r_\mathrm{mst}$ at this value of charge, we obtain a \textit{saddle point} (marginally stable) photon sphere in the CGRN metric. 
A similar merger occurs in the CGS metric ($D_g^2=0$) when $\gamma\rightarrow\pm\infty$~\cite{turner2020}. This phenomenon of the two photon spheres~\eqref{eq:photonspheres} merging to produce the marginally stable photon sphere~\eqref{eq:saddle} is shown in figure~\ref{fig:V_eff} for both $\beta\,\gamma>0$ (figure~\ref{subfig:V_eff_bg+}) and 
$\beta\,\gamma<0$ (figure~\ref{subfig:V_eff_bg-}). 

We note that these saddle point photon spheres are effectively another type of unstable photon sphere. 
Stable rosette orbits require two potential walls that confine the orbiting null particle between radial turning points inside and outside the circular orbit; saddle points have a confining potential on one side, and hence do not allow for actual stable orbits. 

For the $\gamma=0$ case, there is a single unstable photon sphere located at 
\begin{equation}\label{eq:photon_sphere_gamma=0}
    \dfrac{r}{\beta}=\dfrac{r_{\text{ust}}}{\beta}= \dfrac{3}{w_0}.
\end{equation}
This sole photon sphere in the $\gamma = 0$ case is unstable, much like the photon sphere in GRS(A)dS metrics. A nonzero charge pushes out this photon sphere towards larger $r$, reaching $r\rightarrow +\infty$ at $D_g^2\rightarrow1$.

\section{Horizons}\label{sec:horizons}
\subsection{Horizon equations and Hawking temperature}

For metrics of the form~\eqref{eq:ds^2}, horizons are found where $g_{rr}\rightarrow\infty$. In metrics such as the CGRN metric, this naturally occurs for where the blackening factor $B(r)$ vanishes. 

We may rewrite the equation for the locations of horizons in terms of our dimensionless $r/\beta$ values in polynomial form. Taking $\Delta_\mathrm{H}=(r/\beta)\,B(r)$ gives us a cubic equation. For $\gamma\neq0$, we have
\begin{equation}\label{eq:CGRN_horizons}
    \Delta_\text{H}^{\gamma\neq0}\equiv-\beta^2\,\kappa \,\left(\dfrac{r}{\beta}\right)^3 +\beta\,\gamma\,\left(\dfrac{r}{\beta}\right)^2+(1-3\,\beta\,\gamma) \left(\dfrac{r}{\beta}\right)-\left((2-3\,\beta\,\gamma) -\dfrac{D_g^2}{3\,\beta\,\gamma}\right)=0 \ ,
\end{equation}
and for $\gamma=0$,
\begin{equation}\label{eq:CGRN_horizons_gamma0}
    \Delta_\text{H}^{\gamma=0}\equiv-\beta^2\kappa \left(\dfrac{r}{\beta}\right)^3 +w_0\left(\frac{r}{\beta}\right)-2=0 \ .
\end{equation}
Notably, much like the CGS metric, the fact that we have cubic equations means that CGRN has a maximum of $3$ horizons. These may in some instances be Cauchy $\text{H}_\mathrm{C}$, event $\text{H}_\mathrm{E}$, and cosmological $\text{H}_\Lambda$ horizons for increasing $r/\beta$, much like in GRRNdS. 

Once horizon radii are obtained through either~\eqref{eq:CGRN_horizons} or~\eqref{eq:CGRN_horizons_gamma0}, we evaluate the associated Hawking temperatures. 
To do this, we calculate the surface gravity $\mathcal{K}$ at a horizon radius $r_\mathrm{H}$~\cite{costa2020}:
\begin{equation}\label{eq:surface_gravity}
    \mathcal{K} = \dfrac{B'(r_\mathrm{H})}{2} \ ,
\end{equation}
and the Hawking temperature
\cite{changyoung2010} is then:
\begin{equation}\label{eq:hawking_temp}
    T_\mathrm{Hawking}=\dfrac{\mathcal{K}}{2\,\uppi} \ .
\end{equation} 
For CG's static nonrotating metric~\eqref{eq:CG_nonrot_metric}, 
\begin{equation}\label{eq:Hawking_CG_nonrot}
    T_\mathrm{Hawking}(r_\mathrm{H}) = \dfrac{-u+v\,r_\mathrm{H}^2-2\,k\,r_\mathrm{H}^3}{4\,\uppi \,r_\mathrm{H}^2} \ . 
\end{equation}
For the $\gamma\neq0$ CGRN metric~\eqref{eq:CGRN_metric_1},
\begin{equation}\label{eq:Hawking_CGRN1}
    T_\mathrm{Hawking}^{\gamma\neq0} (r_\mathrm{H})=\dfrac{\beta(2-3\,\beta\,\gamma)-\dfrac{D_g^2}{3\,\gamma}+\gamma\, r_\mathrm{H}^2 -2\,\kappa \,r_\mathrm{H}^3}{4\,\uppi \,r_\mathrm{H}^2} \ .
\end{equation}
For the $\gamma=0$ CGRN metric~\eqref{eq:CGRN_metric_2},
\begin{equation}\label{eq:Hawking_CGRN2}
    T_\mathrm{Hawking}^{\gamma=0}(r_\mathrm{H})=\dfrac{2\,\beta - 2\,\kappa\, r_\mathrm{H}^3}{4\,\uppi \,r_\mathrm{H}^2}\ .
\end{equation} 

Note that the negative surface gravity and Hawking temperature at Cauchy $\mathrm{H_C}$ and cosmological $\mathrm{H_\Lambda}$ horizons, where $B'(r)<0$, imply that Hawking radiation is emitted inward to the T\,regions enclosed by these horizons.
At extremal limits, discussed below, two horizons coincide and the Hawking temperature vanishes. 

\subsection{Extremal limits}\label{sec:extr_lims}

We now consider the extremal limits of the spacetime giving the extremal charge $\mathrm{H_{CE}}$, nested limit $\mathrm{H_{EC}}$, Nariai $\mathrm{H_{E\Lambda}}$, and triple limit $\mathrm{H_{TL}}$ horizons. 
Such limits occur where two horizons merge,
and thus where both $B(r)=0$ and $B'(r)=0.$
Note that these two constraints
imply also that $(r^n\,B)'=0$, for any power $n$.
Thus an equivalent pair of constraints
is $\Delta_{\rm H}\equiv(r/\beta)\,B=0$, as in
our horizon equations~\eqref{eq:CGRN_horizons} and~\eqref{eq:CGRN_horizons_gamma0},
and 
$\rmd V_{\text{eff}}/\rmd r =(B/r^2)'= 0$, as 
in our photon sphere constraints in~\eqref{eq:dV} and~\eqref{eq:dVgammazero}.

Accordingly, we find the extremal horizons $\mathrm{H_{CE}}$, $\mathrm{H_{EC}}$, $\mathrm{H_{E\Lambda}}$, and $\mathrm{H_{TL}}$ by simultaneously solving $\Delta_{\mathrm{H}} = 0$ and $\rmd V_{\text{eff}}/\rmd r = 0$. 
Since photon spheres occur where $\rmd V_{\text{eff}}/\rmd r = 0$, as in~\eqref{eq:dV} and~\eqref{eq:dVgammazero}, the extremal limits occur where a horizon coincides with a photon sphere. 
We are thus able to solve for the CG metric parameters $(\beta\,\gamma, \beta^2\,\kappa, D_g)$ corresponding to these extremal limits. 

First dealing with the $\gamma \neq0$ case, substituting $r_{\text{st}}/\beta$~\eqref{eq:photonspheres} in $\Delta^{\gamma \neq 0}_{\mathrm{H}} = 0$~\eqref{eq:CGRN_horizons} gives us an extremal limit at 
\begin{equation}\label{eq: kappa inner photon}
     \left(\beta^2 \kappa\right)^{\gamma\neq0}_{\text{CE}}
    = \frac{\,(\beta\,\gamma)^2 \left[3\,\beta\,\gamma \left(3\,\beta\,\gamma -3w_0 - 2\,
    \right) + (2w_0+1)(w_0+1)\right]}{3\left(3\,\beta\,\gamma - w_0 - 1\right)^3}.
\end{equation}
This corresponds to the extremal charge limit, where an interior Cauchy horizon $\mathrm{H}_\mathrm{C}$ and exterior event horizon $\mathrm{H}_\mathrm{E}$ merge to form $\mathrm{H}_\mathrm{CE}$. 

Now, for the unstable photon sphere at $r_{\text{ust}}/\beta$~\eqref{eq:photonspheres}, we similarly substitute this in to $\Delta^{\gamma \neq 0}_\mathrm{H} = 0$~\eqref{eq:CGRN_horizons} to find 
\begin{equation}\label{eq: kappa outer photon}
    \left(\beta^2 \kappa\right)^{\gamma\neq0}_{\text{EC/E$\Lambda$}}
    =  \frac{\,(\beta\,\gamma)^2 \left[3\,\beta\,\gamma \left(3\,\beta\,\gamma + 3\,w_0 - 2\right) + (2w_0 - 1)(w_0 - 1)\right]}{3\left(3\,\beta\,\gamma + w_0 - 1\right)^3}.
\end{equation}
This in fact describes two limits depending on the sign of $\beta^2\kappa$. When $\beta^2 \kappa < 0$, this corresponds to the nested limit $\mathrm{H}_\mathrm{EC}$ where an interior event horizon $\mathrm{H}_\mathrm{E}$ coincides with an exterior Cauchy horizon $\mathrm{H}_\mathrm{C}$. When $\beta^2 \kappa \geq 0$, this describes the Nariai limit, where an event horizon $\mathrm{H}_\mathrm{E}$ and cosmological horizon $\mathrm{H}_\mathrm{\Lambda}$ coalesce to form $\mathrm{H}_{\mathrm{E}\Lambda}$. Equations~\eqref{eq: kappa inner photon} and~\eqref{eq: kappa outer photon} naturally reduce to the relevant expressions for the CGS metric when $D_g=0$~\cite{turner2020}. 

When $D_g^2 = 1$, as discussed earlier, the equations for stable and unstable photon spheres~\eqref{eq: r photon} coincide to give a marginally stable \textit{saddle point} photon sphere at $r_\mathrm{mst}/\beta$~\eqref{eq:saddle}. Substituting~\eqref{eq:saddle} into~\eqref{eq:CGRN_horizons} corresponds to the extremal triple limit $\mathrm{H}_{\mathrm{TL}}$ with three horizons merging. We write this limit as
\begin{equation}\label{eq:triple limit}
    \left(\beta^2 \,\kappa\right)^{\gamma\neq0}_{\text{TL}} = \frac{(\beta\,\gamma)^2}{9\,\beta\,\gamma - 3}.
\end{equation}

For the $\gamma=0$ metric, we put $r_{\text{ust}}/\beta$ from~\eqref{eq:photon_sphere_gamma=0} into $\Delta^{\gamma = 0}_\mathrm{H} = 0$~\eqref{eq:CGRN_horizons_gamma0}. Thus, the unstable photon sphere intersects a horizon when 
\begin{equation}\label{eq: gamma=0 horizon + photon}
    \left(\beta^2 \,\kappa\right)^{\gamma=0}_{\text{E}\Lambda}= \dfrac{w_0^{3}}{27}. 
\end{equation}
This, similarly to~\eqref{eq: kappa outer photon}, defines the Nariai limit $\mathrm{H}_{\mathrm{E}\Lambda}$ in the $\gamma=0$ case. It also reduces to the relevant limit~\cite{turner2020} of the charge-free CGS metric~\eqref{eq:MK_metric} when we take $D_g = 0$.

\subsection{Horizon structures in CGRN spacetimes}

\begin{figure}
    \centering
    \begin{subfigure}[b]{0.99\textwidth} 
        \includegraphics[width=\textwidth]{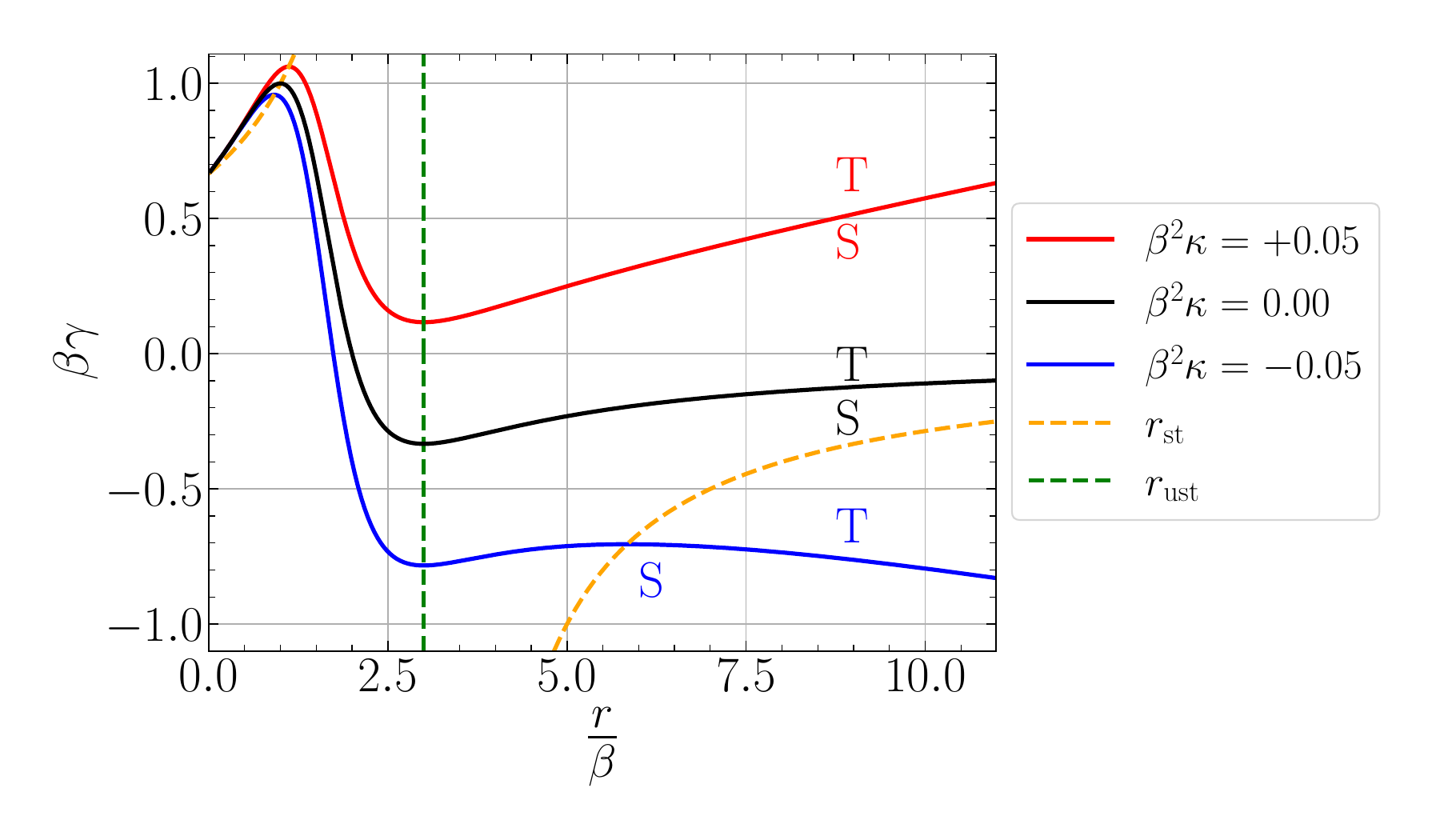} 
        \caption[]%
        {$D_g^2 = 0$ (CGS).} 
        \label{subfig:CGS_structure}
    \end{subfigure}
    \hfill
    \begin{subfigure}[b]{0.99\textwidth}  
        \includegraphics[width=\textwidth]{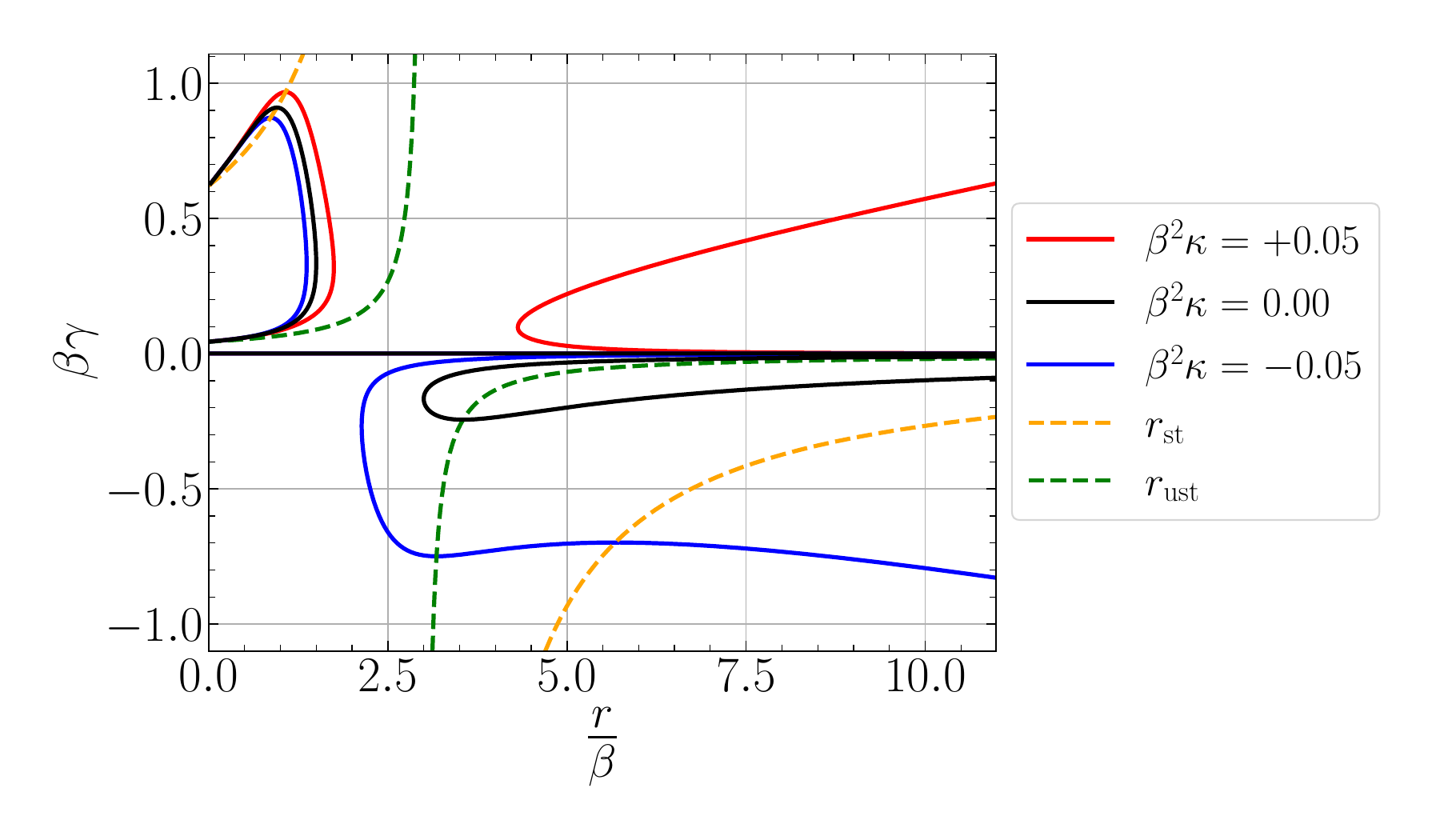}
        \caption[]%
        {$D_g^2 = 0.25$.} 
        \label{subfig:CGRN_structure}
    \end{subfigure}
    \caption{Horizon plots (a) for the CG Schwarzschild metric~\eqref{eq:MK_metric} 
    and
    (b) for the $\gamma\neq0$ CG Reissner-Nordstr\"{o}m metric~\eqref{eq:CGRN_metric_1}.
    In both cases the horizon radii are found as roots of the cubic equation~\eqref{eq:CGRN_horizons} for different values of $\beta\,\gamma$ and $\beta^2\,\kappa$. 
    Solid curves give the horizon radii and dashed ones give photon sphere radii from~\eqref{eq:photonspheres}. 
    We show the full curves for photon sphere solutions, but note that photon spheres cannot exist in spacelike $\mathrm{S}^-$ or $\mathrm{S}^+$ regions, where $B(r)<0$.
    Thus the photon spheres occur above but not below the extremal horizon limits where 2 horizons merge.} 
    \label{fig:D = 0.25 Structure} 
\end{figure}

The radial structure of CGRN spacetimes can be understood by looking at \textit{horizon plots}. 
Figure~\ref{fig:D = 0.25 Structure} shows horizon structures for the CGRN ($\gamma\neq0$) metric~\eqref{eq:CGRN_metric_1} for $\beta^2\,\kappa=0$ and $\pm0.05$. Figure~\ref{subfig:CGS_structure} shows horizons for $D_g^2=0$ (CGS) while figure~\ref{subfig:CGRN_structure} is for $D_g^2=0.25$.
The solid curves give the locations of horizons~\eqref{eq:CGRN_horizons}, and the dashed curves the locations of photon spheres~\eqref{eq: r photon}.
Each horizontal slice corresponds to the radial structure
of the CGRN spacetime with the corresponding value of $\beta\,\gamma$. 
For example, in figure~\ref{subfig:CGS_structure}
 the horizontal slice at $\beta\gamma=0$ crosses the black ($\beta^2\kappa=0$) horizon curve at the Schwarzschild event horizon radius $r=2\,\beta$ and 
the green dashed line at the unstable photon sphere radius $r=3\,\beta$.
More generally, the horizon curves in figure~\ref{subfig:CGS_structure} separate timelike T\,regions above and spacelike S\,regions below. 
Extrema of the horizon curves correspond to extremal limits where two horizons coincide.
The local maxima correspond to extremal horizons $\mathrm{H}_\mathrm{CE}$~\eqref{eq: kappa inner photon}, and local minima to nested limit horizons $\mathrm{H}_\mathrm{EC}$ when $\beta^2\,\kappa<0$ and Nariai horizons  $\mathrm{H}_{\mathrm{E}\Lambda}$ when $\beta^2 \kappa \geq 0$~\eqref{eq: kappa outer photon}~\cite{castro2023}. 
Note that the photon sphere curves intersect with extrema of the horizon curves, and that photon spheres occur only in T\,regions above the horizon curves, where $B(r)>0$, and not in the S\,regions below the horizon curves, where $B(r)<0$.


\begin{figure}[h!] 
    \centering
    \begin{subfigure}[b]{0.7\textwidth} 
        \includegraphics[width=\textwidth]{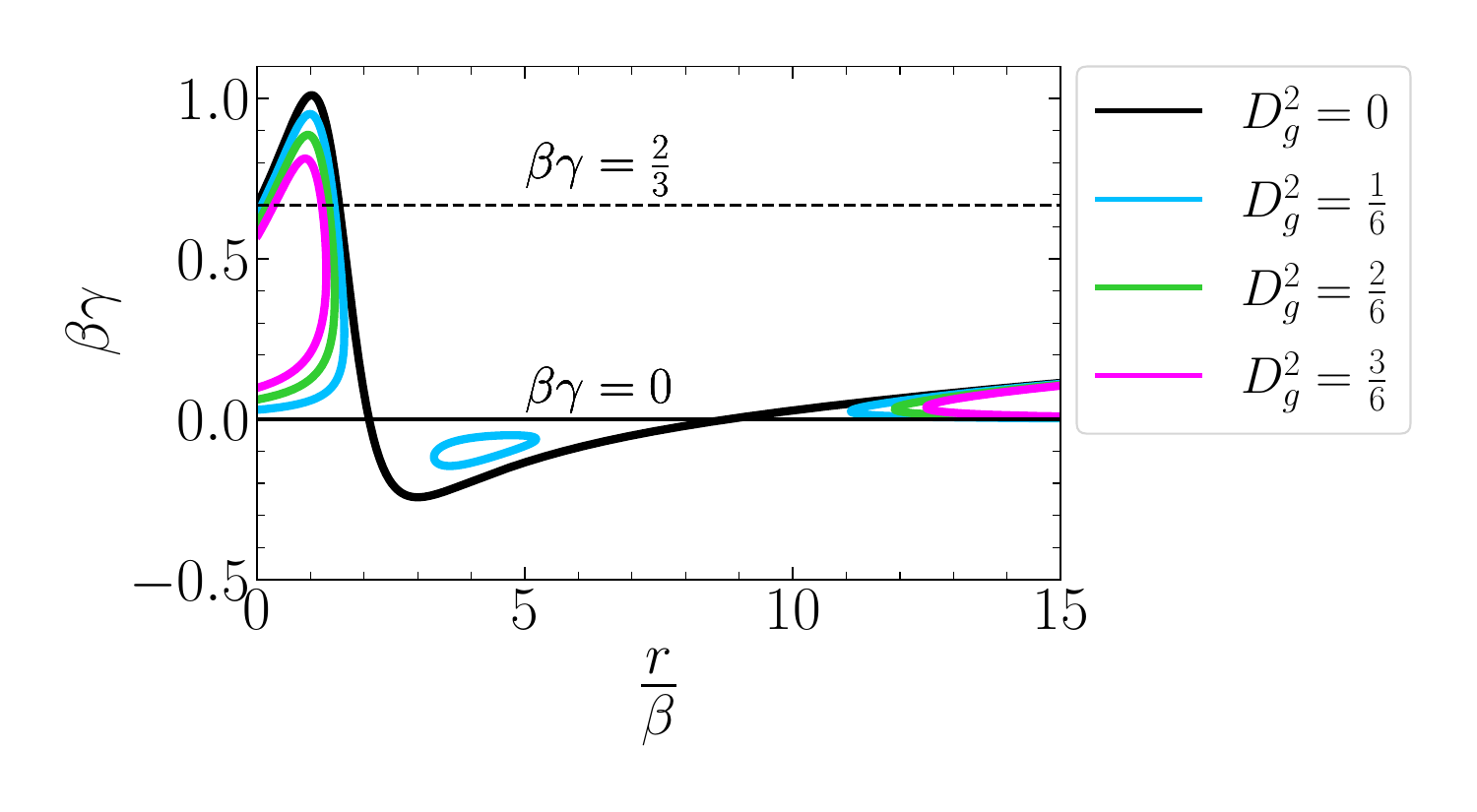} 
        \caption[]%
        {$\beta^2 \,\kappa = 0.01$.}  
        \label{fig: mult_kappa 0.01}
    \end{subfigure}
    \hfill
    \begin{subfigure}[b]{0.7\textwidth}  
        \includegraphics[width=\textwidth]{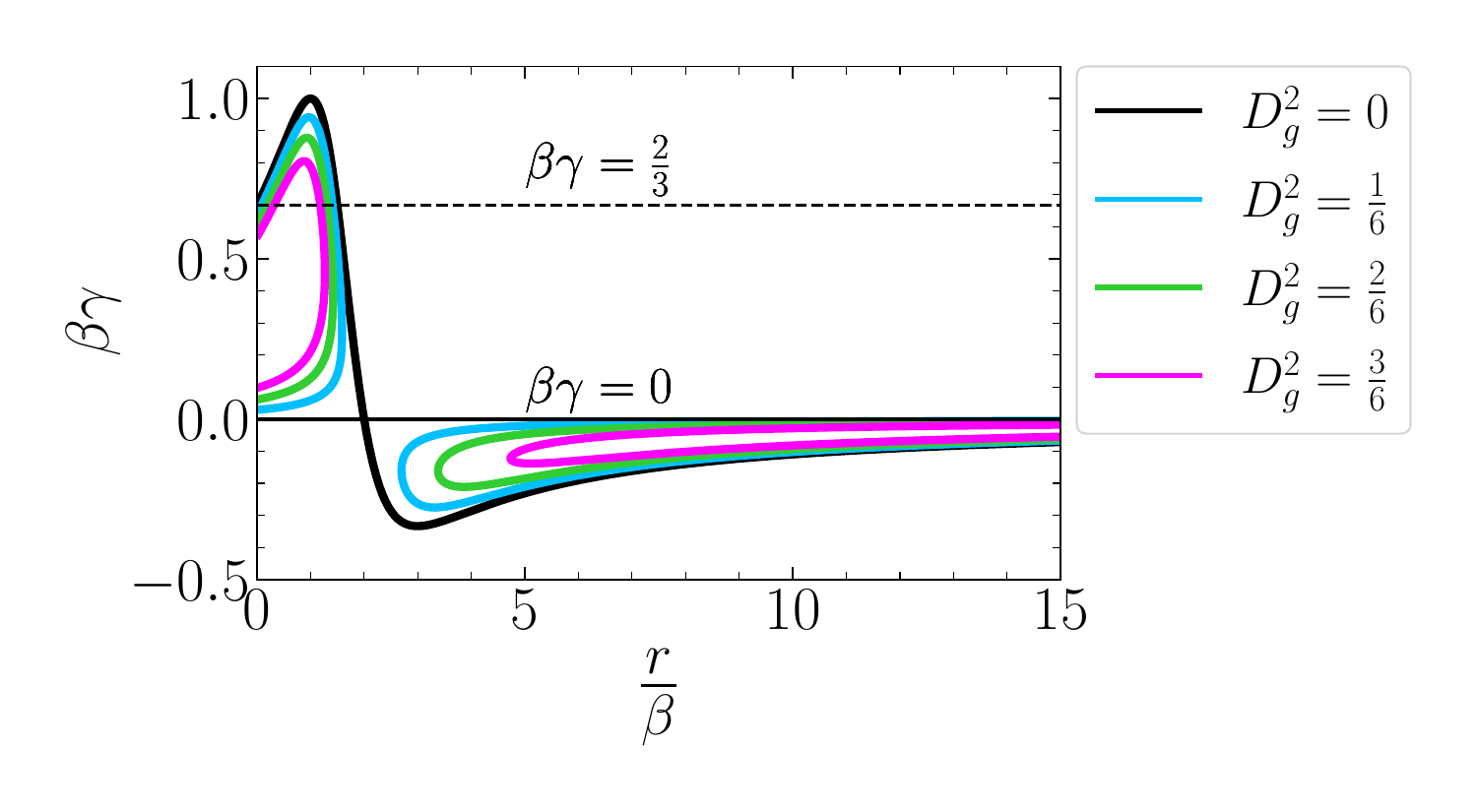}
        \caption[]%
        {$\beta^2 \,\kappa = 0$.} 
        \label{fig: mult_kappa 0}
    \end{subfigure}
    \hfill
    \begin{subfigure}[b]{0.7\textwidth} 
        \includegraphics[width=\textwidth]{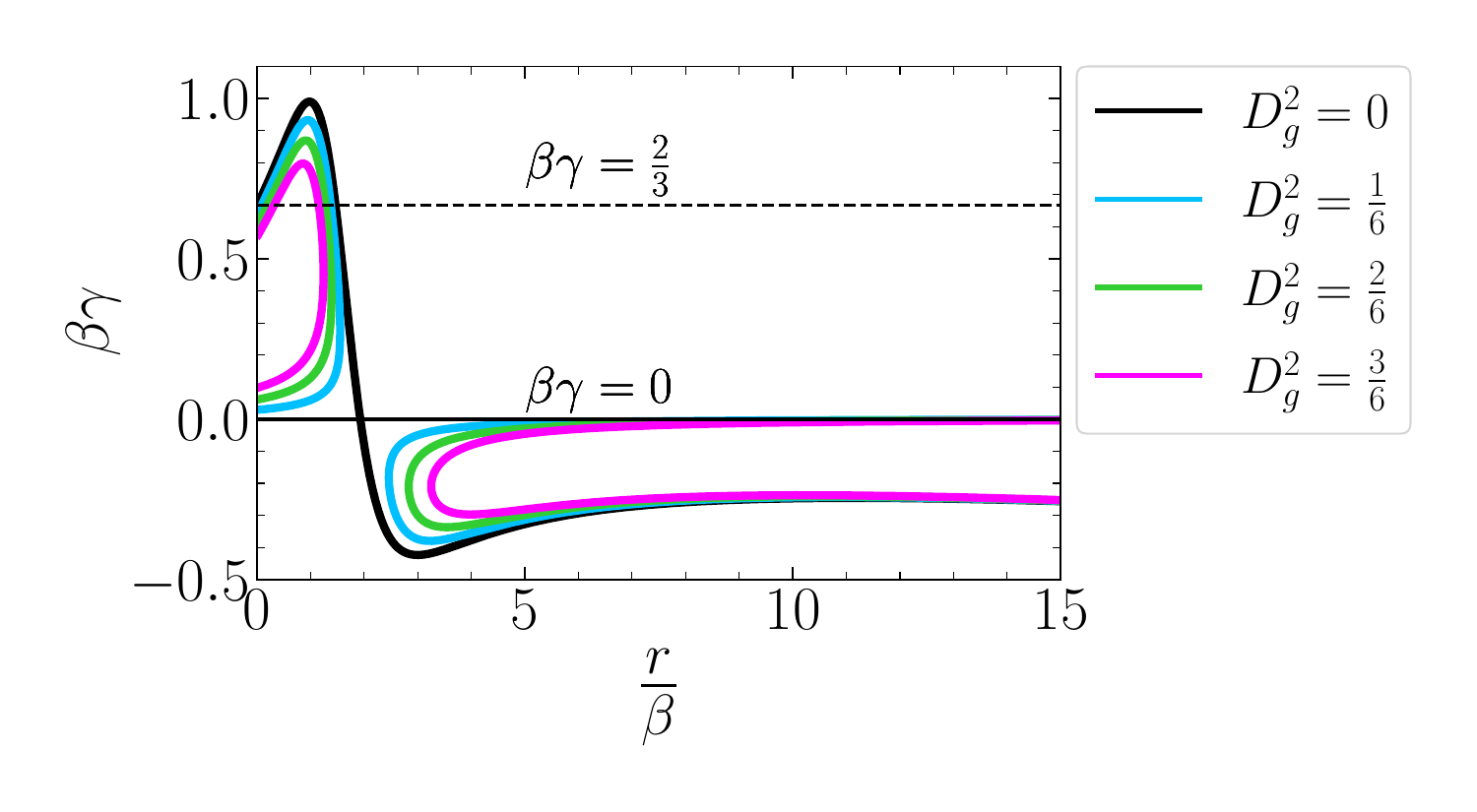}
        \caption[]%
        {$\beta^2\, \kappa = -0.01$.}
        \label{fig: mult_kappa -0.01}
    \end{subfigure}
    \caption[]
    {Horizon curves as in figure~\ref{fig:D = 0.25 Structure}
    but for $\beta^2\,\kappa=0$ and $\pm0.01$. Black, cyan, lime, and magenta correspond to $\Delta_\mathrm{H}^{\gamma\neq0}=0$ for $D_g^2=0$ 
    (CGS), $1/6$, $2/6$, and $3/6$, respectively. The discontinuities at $\beta\,\gamma=0$ are noted as thin black horizontal lines on each subplot for clarity.} 
    \label{fig:MK_vs_CGRN} 
\end{figure} 

As expected, the horizon plot in figure~\ref{subfig:CGRN_structure} for the CGRN ($\gamma\neq0$) metric closely resembles figure~\ref{subfig:CGS_structure}, which represents the CGS metric. Naturally, figure~\ref{subfig:CGS_structure} is in good agreement with the horizon plot presented in~\cite{turner2020}.
The most visible effect of introducing the dyonic charge $D_g$ is the creation of a discontinuity at $\beta\,\gamma=0$.
This difference is also clearly seen in figure~\ref{fig:MK_vs_CGRN},
where the black horizon curves for $D_g^2=0$ 
cross $\beta\,\gamma=0$ at $r/\beta\sim2$ while the
coloured curves for $D_g^2>0$ do not.
This singular behaviour aligns with~\eqref{eq:CGRN_metric_1} giving $B_{\text{CGRN}}^{\gamma \neq 0}(r)$ being ill-defined at $\gamma=0$.

We can use figure~\ref{fig:MK_vs_CGRN} to more clearly understand the effect of charge $D_g^2$ on the spacetime structure. Specifically, when $\beta\gamma>0$, the charge in the $1/r$ term has a positive $D_g^2/3\beta\gamma>0$ contribution. The charge term being positive is akin to the GRRN metric in standard theory where the charge term going as $1/r^2$ is always positive. Similarly, positive values of $\beta\gamma$ correspondingly lead the term $D_g^2/3\beta\gamma$ in CGRN to produce an analogous mechanism, where as charge $D_g^2$ is increased, event horizons are pulled inwards and Cauchy/cosmological horizons are pushed outwards. 

Now, when $\beta\gamma<0$, we have the opposite effect as $D_g^2/3\beta\gamma<0$. In such cases, as clearly observed in figures~\ref{fig: mult_kappa 0} and~\ref{fig: mult_kappa -0.01}, event horizons are pushed outwards, while it is the Cauchy/cosmological horizons that are pulled inwards. This is in direct contrast to the role of charge in standard second-order GR. It can be seen from this that, unlike in the GRRN metric where increasing charge always pulls together horizons, the role of charge depends upon the sign of $\beta\gamma$ in CG. 

It can also be seen that while charge and spin always lead to similar \textit{repulsive} effects on spacetimes in GR, the same is not true in CG. CG's uncharged and rotating Kerr (CGK) metric has spin $a$ always acting repulsively for $\beta > 0$. This leads to CGK black hole spacetimes always having a $\mathrm{T}$ region surrounding its ring singularity~\cite{yulo2025}. Meanwhile, as we find here, CGRN black hole spacetimes can have both spacelike and timelike singularities.

\begin{figure}
    \centering
    \includegraphics[width=0.99\linewidth]{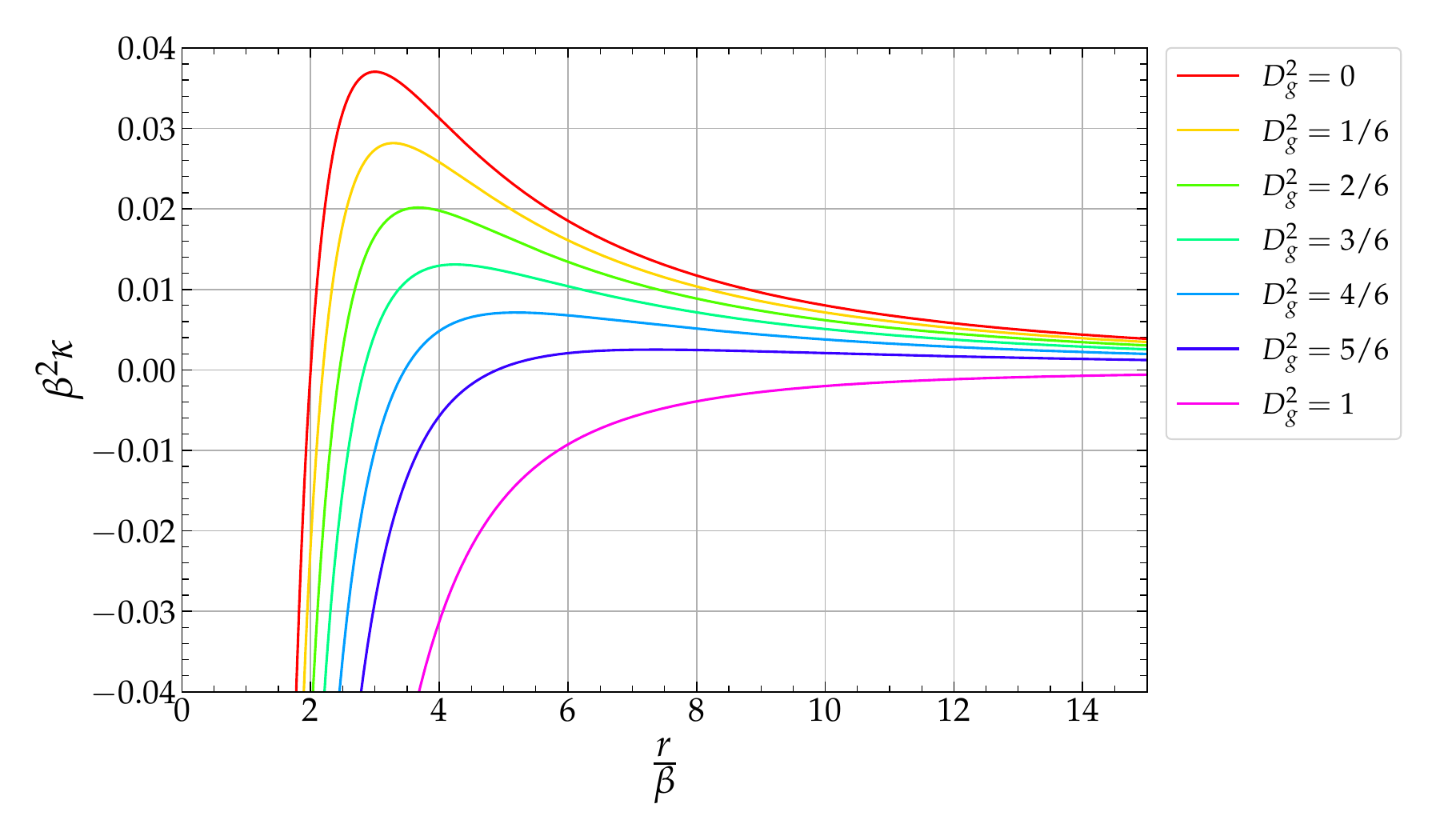}
    \caption{Horizon plots for $\gamma=0$ CGRN spacetimes~\eqref{eq:CGRN_horizons_gamma0}, for various values of $D_g^2$.}
    \label{fig:gamma=0 horizons}
\end{figure}

In contrast to the $\gamma\neq0$ metric~\eqref{eq:CGRN_metric_1}, due to the lack of the linear $\gamma r$ term, the $\gamma=0$ metric~\eqref{eq:CGRN_metric_2} is effectively a GRS(A)dS metric with a modified constant term. Figure~\ref{fig:gamma=0 horizons}, the corresponding horizon plot for~\eqref{eq:CGRN_horizons_gamma0}, can be read in much the same way as figure~\ref{fig:D = 0.25 Structure} or figure~\ref{fig:MK_vs_CGRN}, except now each horizontal slice is associated with a given value of $\beta^2\kappa$ instead of $\beta\,\gamma$. 

Due to the $1/r$ term dominating small $r$ behavior in $B^{\gamma = 0}_{\text{CGRN}}(r)$~\eqref{eq:CGRN_metric_2} associated with the mass $\beta$ always being positive, we never see a Cauchy horizon $\mathrm{H_C}$ interior to an event horizon $\mathrm{H_E}$. Equivalently, the constant term in $\Delta_\mathrm{H}^{\gamma = 0}$ (\ref{eq:CGRN_horizons_gamma0}) is always negative. 

Whether the metric then has a cosmological horizon is dictated by the sign of $\beta^2\kappa$. If $\beta^2\kappa>0$, then there is a cosmological $\mathrm{H}_\Lambda$ horizon if $w_0=(1-D_g^2)^{1/2}>0$. The notable exception is for the maximum dyonic charge $D_g^2=1$. In this case, only $\beta^2\kappa<0$ admits timelike T regions ($B(r)>0$), due to the blackening factor~\eqref{eq:CGRN_metric_2} becoming
\begin{equation}
    B^{\gamma = 0}_{\text{CGRN}}\left(r;D_g^2=1\right)=-\dfrac{2\,\beta}{r}-\kappa \,r^2.
\end{equation}

\section{Classification of CGRN spacetime domains}\label{sec:spacetime_domains}

As has been done for the 
CG Schwarzschild~\cite{turner2020} and CG Kerr~\cite{yulo2025} metrics, we now classify the remarkable variety of spacetimes obtainable in the CGRN metric.
Restricting our study to positive mass $\beta > 0$ and attractive gravity $\alpha_g < 0$, we map out and discuss the spacetime structures that occur as functions of dimensionless metric parameters $\beta\,\gamma$, $\beta^2\,\kappa$, and dyonic charge $D_g^2$. 

Table~\ref{tab:spacetimes} summarises the spacetime structures we find for both CGS ($D_g^2=0$) and CGRN ($D_g^2 > 0$) metrics. 
The CGS metric admits thirteen
distinct spacetime configurations. 
The CGRN metric boasts twenty distinct varieties, seventeen for $\gamma\neq0$ and three for $\gamma=0$. 
Eleven have black holes, with an event horizon $\mathrm{H_E}$.

For straightforward comparison to the corresponding discussion for CGS in~\cite{turner2020}, we include the notation used there denoting the radial order of horizons and photon spheres.
S and U denote stable and unstable photon spheres~\footnote{We correct some of the classifications of CGS from~\cite{turner2020}, where the spacetime domain labels indicate photon spheres within spacelike $\mathrm{S}^-$ and $\mathrm{S^+}$ regions, which are unphysical because test particles in these S regions must move in $r$.}. 
I denotes a ``one-way-in'' horizon with $B'(r)>0$, 
which is an event horizon $\mathrm{H_E}$ with an interior S$^-$ and an exterior T region.
O denotes a horizon with $B'(r)<0$,
which may be a Cauchy $\mathrm{H_C}$ (T inside S$^-$)
or cosmological $\mathrm{H_\Lambda}$ (T inside S$^+$) horizon.
E stands for empty spacetimes with no horizons or photon spheres. 
For CGRN we need an extra marker not present in~\cite{turner2020}: we use M for marginally stable $r_\mathrm{mst}$ photon spheres.

\begin{figure*}
    \centering
    \includegraphics[width=0.999\linewidth]{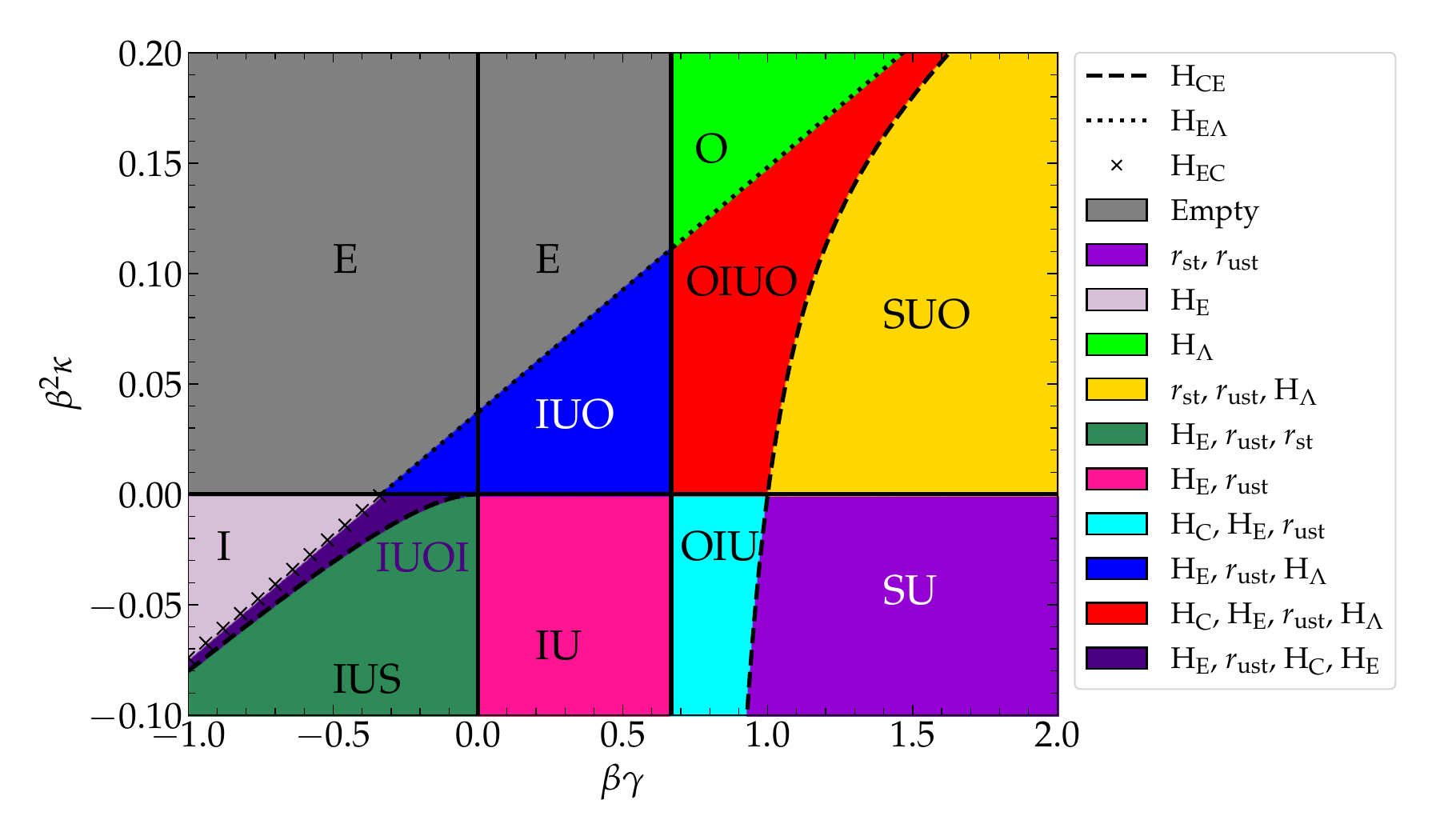}
    \caption{A dimensionless parameter map of spacetimes in the chargeless CG Schwarzschild metric~\eqref{eq:MK_metric}. Refer to table~\ref{tab:spacetimes} for the various classifications. The regions in the legend are ordered in terms of increasing number of horizons.} 
    \label{fig:map_CGS}
\end{figure*}

We map the different spacetimes given by the CGRN metric for various values of $D_g^2$, by creating dimensionless parameter maps of $\beta\,\gamma$ versus $\beta^2\kappa$, as has been done for the CGS metric in~\cite{turner2020} and for the CG\,Kerr metric in~\cite{yulo2025}.
In figures~\ref{fig:map_CGS},~\ref{fig:maps_CGRN_1},~\ref{fig:maps_CGRN_2}, and~\ref{fig:map_gamma=0}, each point on the map corresponds to an entire spacetime structure for a given combination of $\beta\,\gamma$ and $\beta^2\,\kappa$, or alternatively, a horizontal slice in the horizon plots presented in figures~\ref{fig:D = 0.25 Structure} and~\ref{fig:MK_vs_CGRN}. The dashed and dotted lines correspond to the extremal limits presented in~\eqref{eq: kappa inner photon} and~\eqref{eq: kappa outer photon}, where the stable and unstable photon spheres collide with a horizon. 

As stated in section~\ref{sec:extr_lims}, these photon sphere borders may be interpreted as markers for extremal limits. For example, in figure~\ref{fig:map_CGS}, taking the transition from red (OIUO) to lime (O), we note that the difference in these two spacetimes is a result of the annihilation of the black hole event horizon $\mathrm{H}_\mathrm{E}$ (I) with the outermost cosmological horizon $\mathrm{H_\Lambda}$ (O), denoting a cosmological Nariai-like limit $\mathrm{H_{E\Lambda}}$. Here, the unstable photon sphere between the event and cosmological horizons enters a spacelike region, and disappears. Therefore, we may also interpret the formation of the Nariai $\mathrm{H}_{\mathrm{E}\Lambda}$ horizon on the dotted line between these two domains, which is to be expected from~\eqref{eq: kappa outer photon}. 

We start off by looking at the structures of CGRN for $\gamma\neq0$ in section~\ref{subsec:gamma_neq0_maps}, and then investigate the $\gamma=0$ metric spacetimes in section~\ref{subsec:gamma_eq0_maps}. 

\begin{table}[]
\centering
\begin{tabular}{cccccccc}
\hline
\multirow{2}{*}{$\beta\,\gamma$} & \multirow{2}{*}{$\beta^2\,\kappa$} & \multirow{2}{*}{Singularity} & \multirow{2}{*}{Physical features} & \multirow{2}{*}{Structure} & Turner & \multirow{2}{*}{$D_g^2$}      \\ 
  &       &        &  &                                   &         notation            &  \\ \hline 
$+$  &  $+$  &  T/NS  & $r_\mathrm{st}$, $r_\mathrm{ust}$, H$_\Lambda$                & T, S$^{+}$                                         & SUO    & $\left[0,1\right)$  \\[3pt] 
$+$  &  $+$  &  T/BH  & H$_\mathrm{C}$, H$_\mathrm{E}$, $r_\mathrm{ust}$, H$_\Lambda$ & T, S$^{-}$, T, S$^{+}$ & OIUO   & $\left[0,1\right)$  \\[3pt] 
$+$  &  $+$  &  S/BH  & H$_\mathrm{E}$, $r_\mathrm{ust}$, H$_\Lambda$ & S$^{-}$, T & IUO   & $\left[0,1\right)$  \\[3pt] 
$+$  &  $+$  &  S/NS  & Empty & S$^{+}$ & E  & $\left[0,1\right)$  \\[3pt] 
$+$  &  $+$  &  T/NS  & H$_\Lambda$ & T, S$^{+}$                                    & O  & $\left[0,+\infty\right)$  \\[3pt] 
$+$  &  $+$  &  T/NS  & $r_\mathrm{mst}$, H$_\Lambda$ & T, S$^+$ & MO  & $1$  \\[3pt] 
\hline
$+$  &  $-$  &  T/NS  & $r_\mathrm{st}$, $r_\mathrm{ust}$ & T & SU  & $\left[0,1\right)$  \\[3pt] 
$+$  &  $-$  &  T/BH  & $\mathrm{H}_\mathrm{C}$, $\mathrm{H}_\mathrm{E}$, $r_\mathrm{ust}$ & T, S$^{-}$, T & OIU  & $\left[0,1\right)$  \\[3pt] 
$+$  &  $-$  &  S/BH  & $\mathrm{H}_\mathrm{E}$, $r_\mathrm{ust}$ & S$^{-}$, T & IU  & $\left[0,1\right)$  \\[3pt] 
$+$  &  $-$  &  T/NS  & Empty & T & E    & $\left(0, +\infty\right)$  \\[3pt] 
$+$  &  $-$  &  T/NS  & $r_\mathrm{mst}$ & T & M  & $1$  \\[3pt] 
\hline
$-$  &  $+$  &  S/BH  & $\mathrm{H}_\mathrm{E}$, $r_\mathrm{ust}$, $\mathrm{H}_\Lambda$ & S$^{-}$, T, S$^{+}$ & IUO  & $\left[0,1\right)$  \\[3pt] 
$-$  &  $+$  &  S/NS  & Empty & S$^{+}$ & E         & $\left[0,+\infty\right)$  \\[3pt] 
\hline
$-$  &  $-$  &  S/BH  & $\mathrm{H}_\mathrm{E}$, $r_\mathrm{ust}$, $r_\mathrm{st}$ & S$^{-}$, T & IUS  & $\left[0,1\right)$  \\[3pt] 
$-$  &  $-$  &  S/BH  & $\mathrm{H}_\mathrm{E}$, $r_\mathrm{ust}$, $\mathrm{H}_\mathrm{C}$, $\mathrm{H}_\mathrm{E}$ & S$^{-}$, T, S$^{-}$, T & IUOI   & $\left[0,1\right)$  \\[3pt] 
$-$  &  $-$  &  S/BH  & $\mathrm{H}_\mathrm{E}$ & S$^{-}$, T & I  & $\left[0,+\infty\right)$  \\[3pt] 
$-$  &  $-$  &  S/BH  & $\mathrm{H}_\mathrm{E}$, $r_\mathrm{mst}$ & S$^-$, T & IM  & $1$  \\[3pt] 
\hline
$0$ & $+$ & S/BH & $\mathrm{H}_\mathrm{E}$, $r_\mathrm{ust}$, $\mathrm{H}_\Lambda$ & S$^-$, T, S$^+$ & IUO & $\left[0,1\right)$  \\[3pt] 
$0$ & $+$ & S/NS & Empty & S$^+$ & E  & $\left[0,1\right]$  \\[3pt] 
$0$ & $-$ & S/BH & $\mathrm{H}_\mathrm{E}$, $r_\mathrm{ust}$ & S$^-$, T & IU  & $\left[0,1\right)$  \\[3pt] \hline 
\end{tabular} 
\caption{Spacetime configurations available in the CG Schwarzschild $(D_g^2 = 0)$ and CG Reissner-Nordstr\"{o}m $(D_g^2 > 0)$ metrics for $\beta>0$. In the \textit{Singularity} column, T/S denote timelike (T: $B(0)\rightarrow+\infty$) or spacelike (S: $B(0)\rightarrow-\infty$) singularities, and BH/NS determine whether these singularities are concealed by an event horizon $\mathrm{H}_\mathrm{E}$ (BH), or whether they are naked (NS). The Turner~\cite{turner2020}  notation indicates the radial order of horizons and photon spheres: I and O denote in and out horizons with $B'>0$ and $B'<0$ respectively, and S and U denote stable and unstable photon spheres~\eqref{eq:photonspheres}. M denotes the \textit{saddle point} (marginally stable) photon sphere~\eqref{eq:saddle} where S and U merge. E denotes an empty spacetime with neither horizons nor physical photon spheres.} 
\label{tab:spacetimes} 
\end{table}

\subsection{\texorpdfstring{$\gamma\neq0$}{gamma neq 0} CGRN spacetimes}\label{subsec:gamma_neq0_maps} 

\begin{figure*}[h!]
    \centering
    \begin{subfigure}[b]{0.999\textwidth}
        \includegraphics[width=\textwidth]{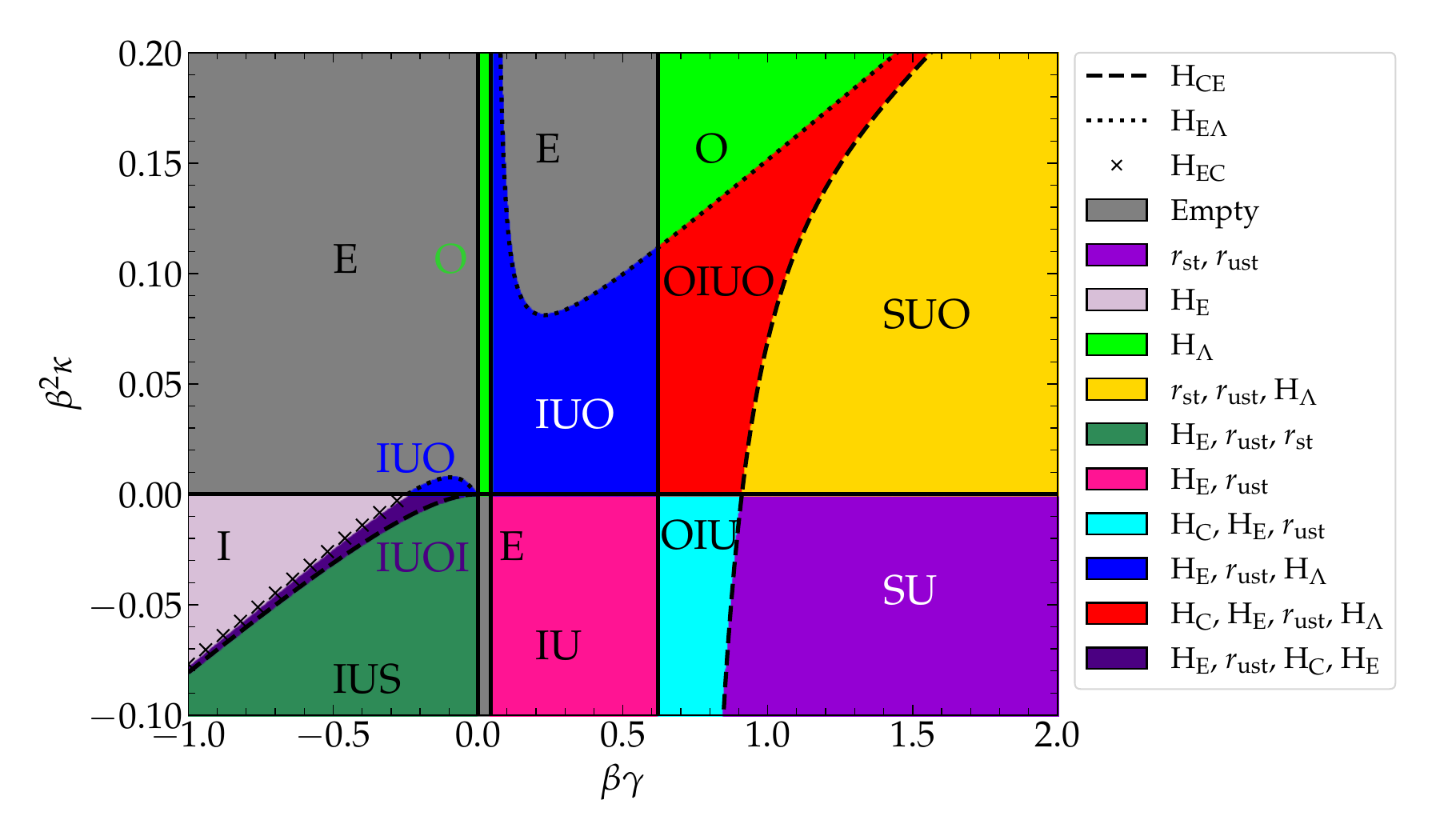}
        \caption[]%
        {$D_g^2=1/4$.}
        \label{fig:map_CGRN_2/3}
    \end{subfigure}
    \hfill
    \begin{subfigure}[b]{0.999\textwidth}
        \includegraphics[width=\textwidth]{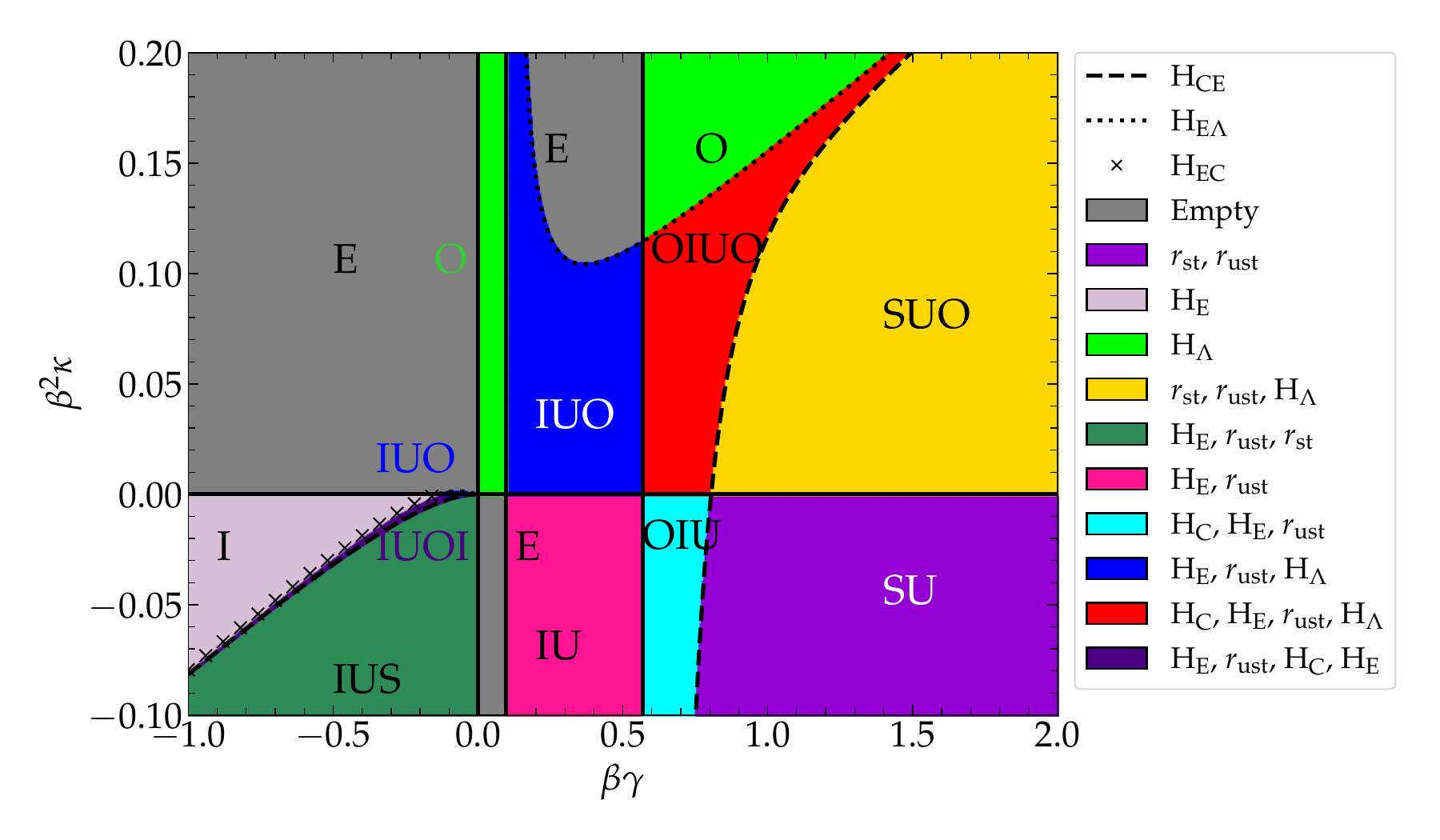}
        \caption[]%
        {$D_g^2=1/2$.}
        \label{fig:map_CGRN_4/3}
    \end{subfigure}
    \caption[]
    {Dimensionless parameter maps of $\gamma\neq0$ CG Reissner-Nordstr\"{o}m spacetimes, for nonzero $D_g^2$ values below the critical $D_g^2=1$. Note that as the charge increases, the extremal limit borders~\eqref{eq: kappa inner photon} and~\eqref{eq: kappa outer photon} approach each other.}
    \label{fig:maps_CGRN_1}
\end{figure*}

Comparing the charge-free $(D_g^2 = 0)$ map of CGS spacetimes in figure~\ref{fig:map_CGS} with CGRN maps for $D_g^2=0.25$ and 0.5
in figure~\ref{fig:maps_CGRN_1}, we see that the dyonic charge introduces significant changes. 
The vertical boundary at $\beta\,\gamma=0$ remains in place but spawns a new vertical boundary that moves to larger $\beta\,\gamma$, opening a gap with two new domains. 
One of these ($\beta^2\,\kappa>0$) has a cosmological horizon $\mathrm{H}_\Lambda$ (O), and the other ($\beta^2\,\kappa>0$) is a completely empty spacetime (E). 
This vertical boundary is of course the asymptote for~\eqref{eq: kappa outer photon}. 
Meanwhile, a second vertical boundary
at $\beta\,\gamma=2/3$ for CGS, corresponding to the stable photon sphere extremal limit
$\mathrm{H_{CE}}$~\eqref{eq: kappa inner photon}, moves towards decreasing $\beta\,\gamma$ as $D_g^2$ increases.
These two vertical boundaries come together at 
$\beta\,\gamma=1/3$ for $D_g^2=1$.

\begin{figure*}[h!] 
    \centering
    \begin{subfigure}[b]{0.999\textwidth} 
        \includegraphics[width=\textwidth]{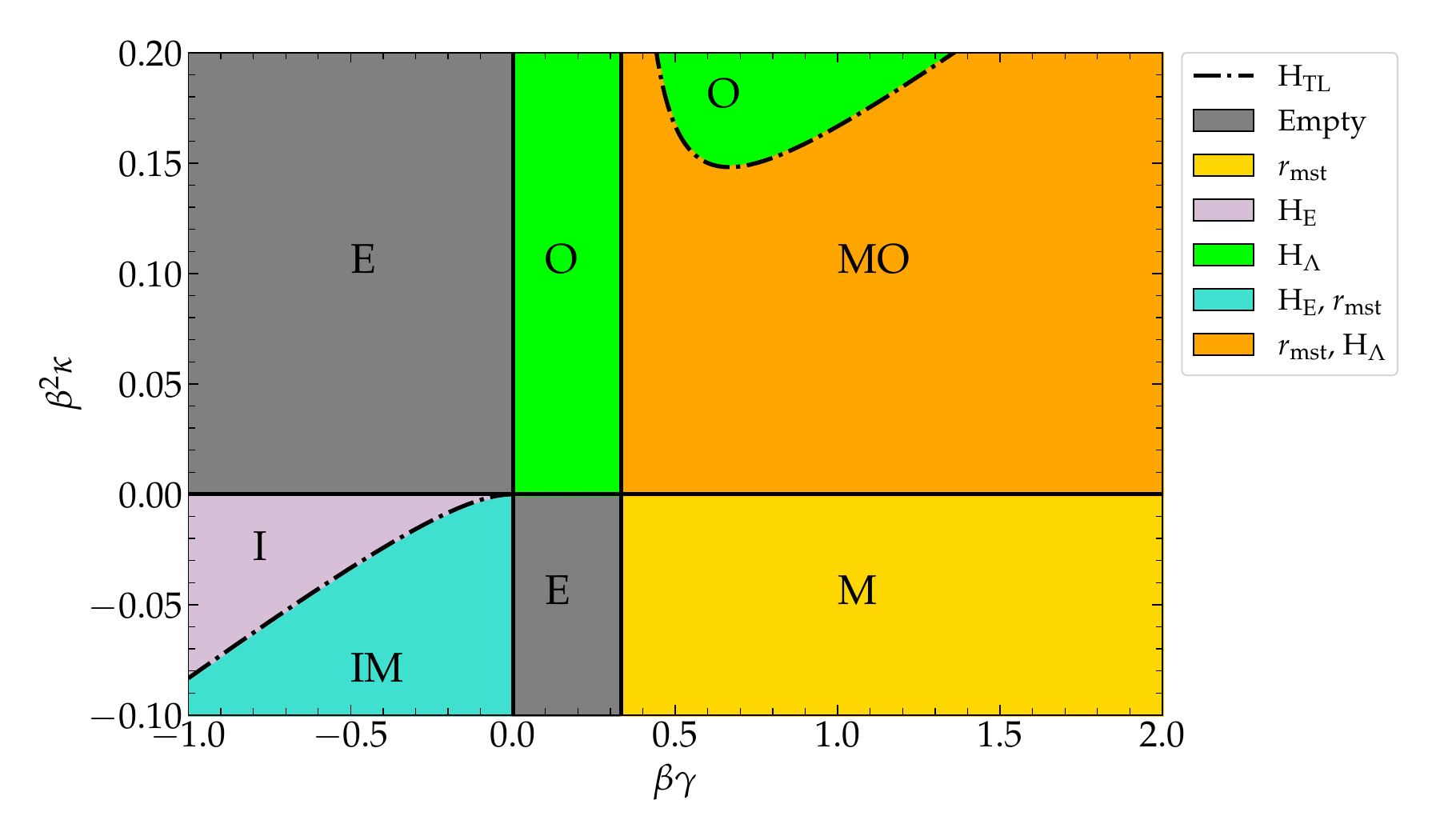} 
        \caption[]%
        {$D_g^2=1$.}    
        \label{fig:map_CGRN_8/3}
    \end{subfigure}
    \hfill
    \begin{subfigure}[b]{0.999\textwidth} 
        \includegraphics[width=\textwidth]{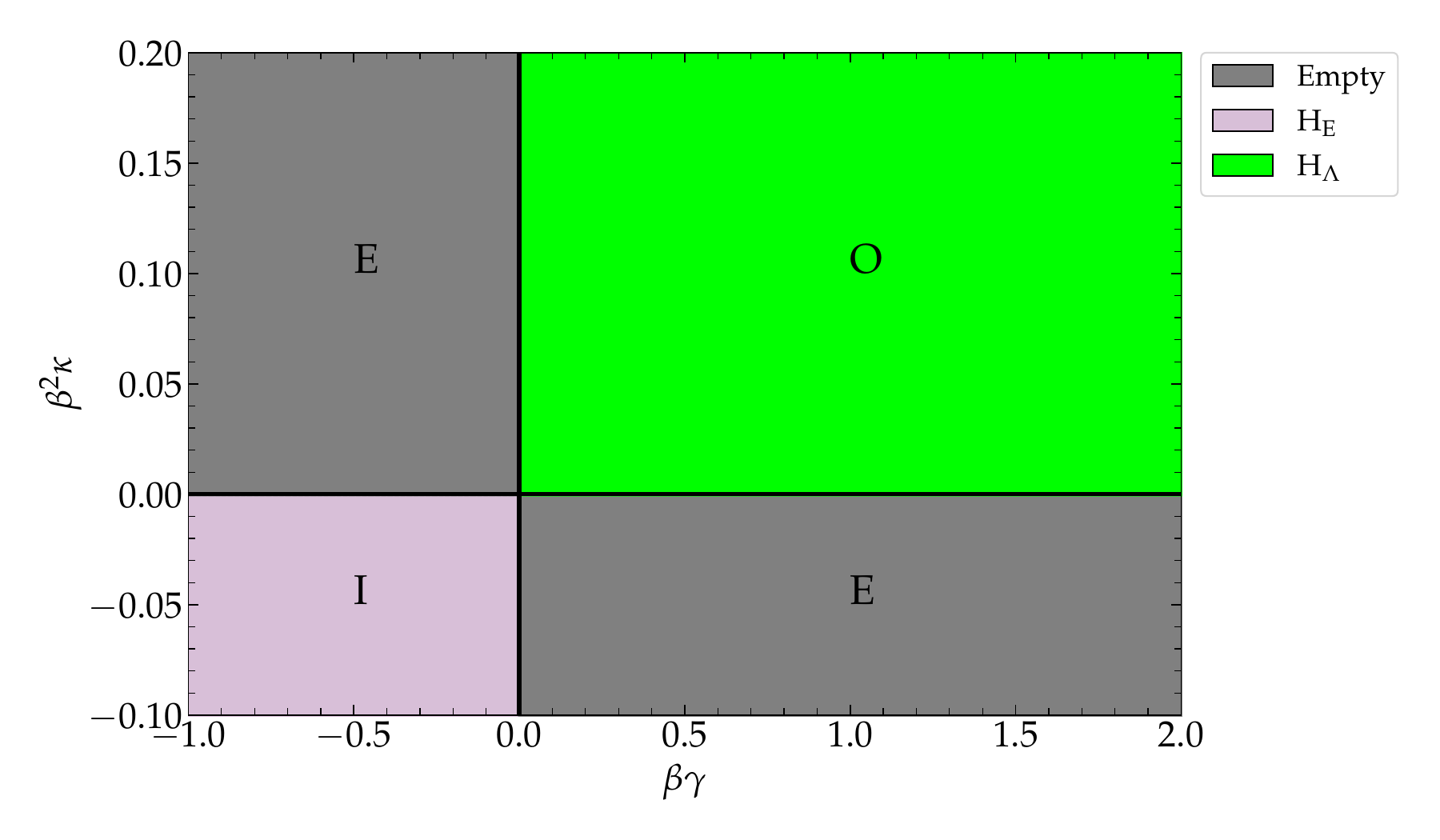} 
        \caption[]%
        {$D_g^2=9/8$.} 
        \label{fig:map_CGRN_9/3} 
    \end{subfigure} 
    \caption[] 
    {Dimensionless parameter maps of $\gamma\neq0$ CG Reissner-Nordstr\"{o}m spacetimes, for nonzero $D_g^2$ values at and above the critical $ D_g^2 = 1$ value. Note the appearance of extremal triple limit horizon $\mathrm{H}_\mathrm{TL}$~\eqref{eq:triple limit} in figure~\ref{fig:map_CGRN_8/3} shown by the dashdotted line, and its consequent disappearance above this limit in figure~\ref{fig:map_CGRN_9/3}.} 
    \label{fig:maps_CGRN_2} 
\end{figure*}

The CGRN map exhibits two further major changes: one at $D_g^2=1$ (figure~\ref{fig:map_CGRN_8/3}), and the other just above this threshold charge (figure~\ref{fig:map_CGRN_9/3}). 
At $D_g^2=1$, the two photon sphere borders~\eqref{eq: kappa inner photon} and~\eqref{eq: kappa outer photon} merge. This corresponds to the stable and unstable photon spheres converging to form
the marginally-stable saddle-point photon sphere $r_\mathrm{mst}$ discussed earlier in section~\ref{sec:null_geodesics}. 
This phenomenon does not occur in CGS unless $\gamma\rightarrow\pm\infty$. 
As~\eqref{eq: kappa inner photon} and~\eqref{eq: kappa outer photon} coincide with $\mathrm{H}_\mathrm{CE}$ and $\mathrm{H}_{\mathrm{E}\Lambda}$ respectively, the threshold charge $D_g^2=1$ gives rise to the extremal triple limit horizon $\mathrm{H_{TL}}$ mentioned in section~\ref{subsec:horizon_types}. 
For example, in GRRNdS spacetimes
($\beta\gamma=0$, $\beta^2\kappa>0$ in figure~\ref{fig:map_CGS}), the Cauchy, event, and cosmological horizons
($\mathrm{H_C}$, $\mathrm{H_E}$ and $\mathrm{H_\Lambda}$) can merge.
The resulting extremal triple limit horizon ($\mathrm{H_{TL}}$) has a near-horizon geometry of Mink$_2\times$S$^2$, and is sometimes referred to as the \textit{ultracold limit}~\cite{castro2023}. 
Noting its formation sequence, we may also refer to this as $\mathrm{H_{CE\Lambda}}$.

\begin{figure} 
    \centering 
    \includegraphics[width=0.9\linewidth]{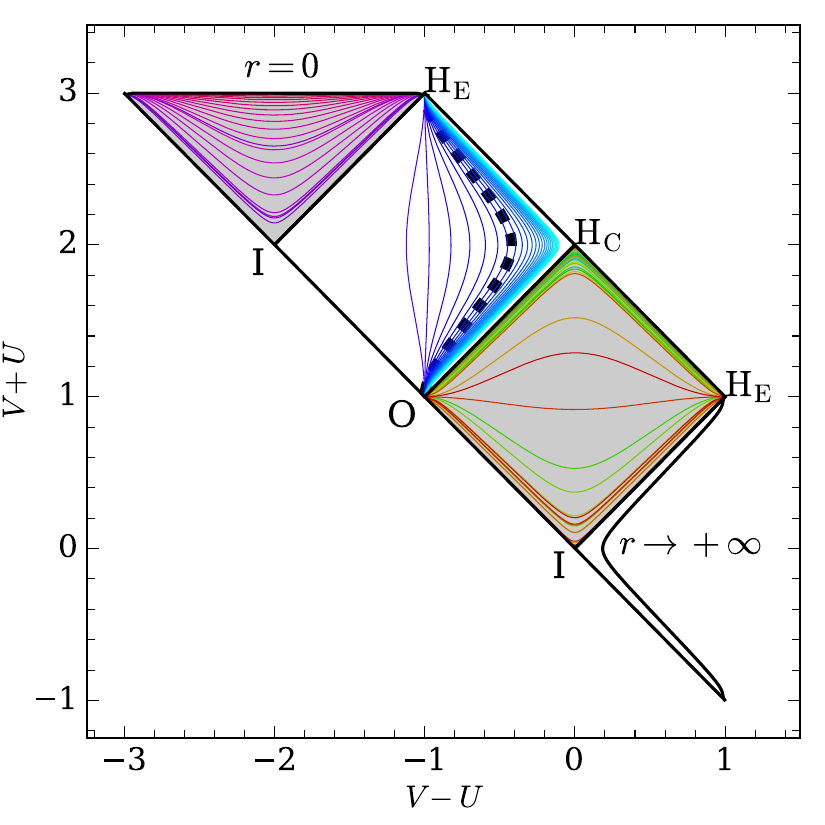} 
    \caption{A conformal Penrose diagram for the ``nested black hole'' structure (IUOI: dark purple in figures~\ref{fig:map_CGS} and~\ref{fig:maps_CGRN_1}) obtainable in the CG Schwarzschild and CG Reissner-Nordstr{\"o}m metrics. The colourful lines are lines of constant radius, and white/grey blocks correspond to T/S$^-$ regions; naturally, the diagonal lines separating these regions correspond to the horizons of the metric. The unstable photon sphere $r_\mathrm{ust}$ is denoted by the black dashed line. \textit{V} and \textit{U} correspond to the compactified null coordinates. Here, $r \rightarrow +\infty$ corresponds to a timelike boundary as in Anti-de Sitter spacetimes. This particular Penrose diagram describes the $\gamma\neq0$ CG Reissner-Nordstr\"{o}m metric~\eqref{eq:CGRN_metric_1} with $\beta\,\gamma=-0.3$, $\beta^2\,\kappa=-0.01$, and $D_g^2=0.25$.} 
    \label{fig:nested_penrose} 
\end{figure}

However, there are other ways in which three horizons can coalesce in CGRN. 
For example, with $\beta\,\gamma < 0$ and $D_g^2<1$, there are CGRN spacetimes with inner $\mathrm{H}_\mathrm{E}$, intermediate $\mathrm{H}_\mathrm{C}$, and outer $\mathrm{H}_\mathrm{E}$ horizons (dark purple IUOI domain in figure~\ref{fig:maps_CGRN_1}), a configuration also present in the CGS map (figure~\ref{fig:map_CGS}). 
This is a \textit{nested} black hole structure akin to a multi-layered GRS metric, for which 
figure~\ref{fig:nested_penrose} shows
the conformal Penrose diagram generated by \texttt{xhorizon}\footnote{\href{https://github.com/xh-diagrams/xhorizon}{https://github.com/xh-diagrams/xhorizon}}~\cite{schindler2018}. 
In CGRN at the threshold charge limit of $D_g^2=1$, these three horizons merge to create a near-horizon geometry distinct from Mink$_2 \times$S$^2$. 
We cannot refer to this triple horizon merger as \textit{ultracold}, as we do not yet understand the thermodynamic response of the metric at this limit. 
For this reason, the generic term (extremal triple limit $\mathrm{H}_\mathrm{TL}$) 
can refer to three coincident horizons regardless of heritage as either
$\mathrm{H_C}$, $\mathrm{H_E}$ and $\mathrm{H_\Lambda}$ merging to form $\mathrm{H_{CE\Lambda}}$, 
or $\mathrm{H_E}$, $\mathrm{H_C}$,
and $\mathrm{H_{E}}$ merging to form
$\mathrm{H_{ECE}}$. 

The two vertical borders 
located at $\beta\,\gamma=0$ and $2/3$
for $D_g^2=0$ (figure~\ref{fig:map_CGS})
are $\kappa\rightarrow\pm\infty$ asymptotes
of the extremal horizons defined by~\eqref{eq: kappa inner photon} and~\eqref{eq: kappa outer photon}. 
These come together as $D_g^2$ increases 
(figure~\ref{fig:maps_CGRN_1})
and converge at $\beta\,\gamma=1/3$ 
to form the $\mathrm{H_{TL}}$ defined by~\eqref{eq:triple limit}
for $D_g^2=1$ (figure~\ref{fig:map_CGRN_8/3}).
This merger effectively 
occludes the three intervening domains:
empty (E: grey), $\mathrm{H_E}, r_\mathrm{ust}, \mathrm{H_\Lambda}$ (IUO: dark blue), and $\mathrm{H_E}, r_\mathrm{ust}$ (IU: dark pink).

$D_g^2>1$ (figure~\ref{fig:map_CGRN_9/3}) removes the borders~\eqref{eq: kappa inner photon} and~\eqref{eq: kappa outer photon} entirely. This is to be expected, considering the absence of photon spheres due to the reality condition from the radical in~\eqref{eq: r photon}. 
$D_g^2>1$ (figure~\ref{fig:map_CGRN_9/3}) removes the borders~\eqref{eq: kappa inner photon} and~\eqref{eq: kappa outer photon} entirely. Note that while such large values of charge are prohibited in the $\gamma=0$ CGRN metric from the reality condition on $w_0$~\eqref{eq:w0}, no such constraint is imposed on the metric when $\gamma\neq0$~\eqref{eq:CGRN_metric_1}. Therefore, if $\gamma\neq0$ we can still obtain spacetimes with $D_g^2>1$, though these will contain no photon spheres. Correspondingly, due to the reality condition from the radical in~\eqref{eq: r photon}, such spacetimes are devoid of extremal horizon limits too. In fact, the only borders are at $\beta\,\gamma=0$ and $\beta^2\kappa=0$. Therefore, CGRN spacetimes with $D_g^2>1$ are either completely empty (E: grey), have a naked timelike singularity with a cosmological $\mathrm{H_\Lambda}$ horizon (O: lime), or resemble a GRSAdS metric sans the unstable photon sphere (I: plum).

\subsection{\texorpdfstring{$\gamma=0$}{gamma = 0} CGRN spacetimes}\label{subsec:gamma_eq0_maps}

\begin{figure}
   \centering
    \includegraphics[width=0.999\linewidth]{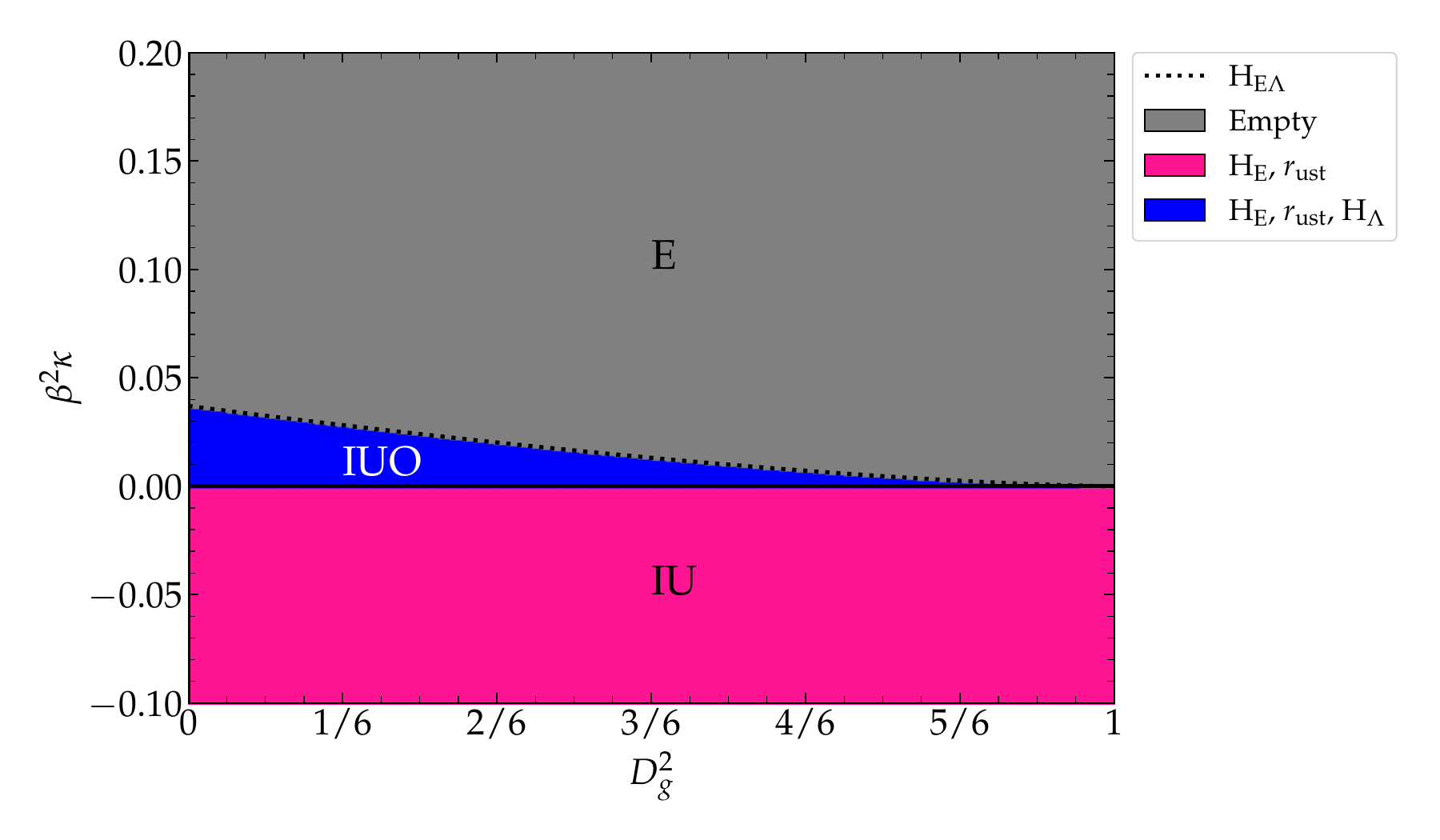}
    \caption{Dimensionless parameter map of $\gamma=0$ CG Reissner-Nordstr\"{o}m spacetimes.
    \label{fig:map_gamma=0}
    }
\end{figure}

For the $\gamma=0$ CGRN metric, our axes for the parameter map in figure~\ref{fig:map_gamma=0} are $\beta^2\kappa$ and $D_g^2$. The number of possible spacetime configurations is significantly reduced compared to the $\gamma\neq0$ cases. 
There are two spacetime configurations with horizons, neither of which has a $\mathrm{H_C}$. 
The first of these (IU: dark pink) has an event horizon $\mathrm{H_E}$ with an exterior unstable photon sphere, much like the GRSAdS metric. 
The second configuration (IUO: dark blue) has both event $\mathrm{H_E}$ and cosmological $\mathrm{H_\Lambda}$ horizons enclosing an unstable photon sphere, as does the GRSdS metric. 
The transition at $\beta^2\,\kappa=0$ is expected because with $\gamma=0$
the blackening factor $B(r)$ at larger radii is dominated by the $-\kappa\,r^2$ term,  and
$\kappa>0$ allows $B(r)=0$ at the $\mathrm{H_\Lambda}$ horizon.
Finally, for $\beta^2\kappa>0$ 
and sufficiently large $D_g^2$,
the two horizons merge to leave an empty
spacetime
with a naked spacelike singularity (E: grey).  
As the dotted line serves as a transition between the $\mathrm{H_E}\rightarrow r_\mathrm{ust} \rightarrow\mathrm{H_\Lambda}$ (IUO) and empty (E) spacetimes, this border represents spacetimes 
with an extremal $\mathrm{H_{E\Lambda}}$ horizon, similar in structure to the cosmological Nariai limit. 

\section{Conclusions and further perspectives}\label{sec:concl}

In the present work, we have 
reviewed the structures of positive-mass Reissner-Nordstr\"{o}m spacetimes in Weyl Conformal Gravity, and defined the various limits and geometries that may be obtained from this metric. Notably, with the inclusion of charge, the varieties of spacetimes are more numerous compared to the Conformal Gravity Schwarzschild metric~\cite{turner2020}. One feature of the Conformal Gravity Reissner-Nordstr\"{o}m metric that distinguishes it from the Conformal Gravity Schwarzschild metric is the possibility of obtaining a marginally stable \textit{saddle point} photon sphere. In the Conformal Gravity Schwarzschild case, this is only possible if $\gamma\rightarrow\pm\infty$. We have also identified that the instances where photon spheres and horizons collide correspond to the extremal limits of the metric.

We identify a critical threshold value for our auxiliary dyonic-gravitational parameter $D_g=\pm{\sqrt{-3(Q^2+P^2)/(8\alpha_g)}}$, which has the constraint $D_g^2 \leq 1$. Any value of $D_g$ falling above this critical threshold of $D_g^2=1$ makes the constant term in the metric become complex when $\gamma=0$. On the other hand, for the $\gamma\neq0$ case, three horizons may collide if $D_g^2=1$, and $D_g^2>1$ removes all photon spheres, yielding either an empty metric, a metric resembling the Schwarzschild-Anti-de Sitter metric without the unstable photon sphere, or a singular metric with a cosmological horizon and de-Sitter curvature. 

We close by outlining several possible extensions of the present work. 
First, elucidating the near-horizon geometries at the extremal limits of the Conformal Gravity Reissner-Nordstr\"{o}m metric may give rise to structures like AdS$_2\times$S$^2$, dS$_2\times$S$^2$, or Mink$_2\times$S$^2$, which are important in certain aspects of quantum gravity and string theory~\cite{castro2023}. Interesting in particular is the near-horizon geometry of this metric at the maximum $D_g$ value, where the stable and unstable photon sphere both collide with the same horizon. As mentioned, this phenomenon is unseen in the Conformal Gravity Schwarzschild metric. Depending on the combination of horizons that collide, we may discover near-horizon geometries unobtainable in general relativity, such as for the nested triple limit involving a Cauchy $\mathrm{H}_\mathrm{C}$ horizon trapped between two event $\mathrm{H}_\mathrm{E}$ horizons. We hope to address this in a future work. 

Second, we note that in the present work, we have not explored spacetime configurations with a negative mass parameter $(\beta<0)$ or a positive gravitational coupling constant $(\alpha_g>0)$. While our justification for excluding this here is so that the metric reduces to familiar attractive gravity in the Newtonian and weak-field limits~\cite{mannheim_2007}, it is nevertheless true that $\beta<0$ and $\alpha_g>0$ are permitted within the theory. We thus encourage future research in this direction as well. 

Another avenue for future work is the application of our current metric analysis techniques to the fourth (electro)vacuum metric of Conformal Gravity, the Kerr-Newman (CGKN) metric, which is analogous to the charged and rotating Kerr-Newman metric of general relativity. 
While its required field equation constraints~\cite{mannheim1991}, thermodynamic properties~\cite{liu2013}, and null geodesics~\cite{fathi2021} have previously been discussed, there is yet to be a comprehensive review of the spacetime structures obtainable in the CGKN metric. 

Finally, we notice that because the external electro-magnetic field sourced by the dyonic charge has a radial pressure that scales
as $r^{-4}$, the CGRN metric admits
a further 2-parameter manifold of solutions that differ in how the dyonic charge affects
the constant, inverse linear, 
and linear terms in the blackening factor $B(r)$, as parametrised by $(p,q)$ in~\eqref{eq:mess}.
Our study adopts the specific choices
of Mannheim \& Kazanas~\cite{mannheim1991},
namely $(p,q)=(0,1)$ for $\gamma\neq0$
and $(p,q)=(1,0)$ for $\gamma=0$.
An even richer variety of spacetimes may be accessible by considering more general choices of $(p,q)$.
Specific values for $(p,q)$ should be determined by matching the exterior CGRN solution to an interior solution,
but these have not yet been studied.

\section*{Acknowledgments}

We thank Philip Mannheim for fruitful discussions regarding Conformal Gravity. We also thank Barak Shoshany for his help with using \texttt{OGRePy}, and Joseph C.\,Schindler for his help on creating Penrose diagrams for custom metrics with \texttt{xhorizon}. We are also grateful to Jack Purllant for his insight into the infinities in our Penrose diagram. We acknowledge the helpful and insightful comments from the three anonymous reviewers of this manuscript.

\section*{References}
\bibliography{Bibliography}

\end{document}